\documentclass[
  reprint,
  aps,
  pra,
  superscriptaddress,
  nofootinbib,
  longbibliography,
  floatfix
]{revtex4-2}

\usepackage[T1]{fontenc}
\usepackage[utf8]{inputenc}
\usepackage{lmodern}
\usepackage{microtype}
\usepackage{amsmath,amssymb,mathtools,bm}
\usepackage{braket}
\usepackage{graphicx}
\usepackage{booktabs}
\usepackage{multirow}
\usepackage{xcolor}
\usepackage{xspace}
\usepackage{tikz}
\usetikzlibrary{calc}
\usepackage{hyperref}

\hypersetup{
  colorlinks=true,
  linkcolor=black,
  citecolor=black,
  urlcolor=black,
  pdftitle={Classical Active-Space Hybrid Quantum Subspace Expansion (CASH-QSE): Quantum Corrections without Remeasuring the Classically Calculable Energy}
}

\newcommand{\Hhat}{\hat H}
\newcommand{\CAS}{\mathrm{CAS}}
\newcommand{\CASH}{CASH-QSE\xspace}

\newcommand{\Span}{\operatorname{span}}

\newcommand{\order}{\mathcal O}
\newcommand{\mel}[3]{\langle #1|#2|#3\rangle}
\newcommand{\vct}[1]{\bm{#1}}
\newcommand{\mEh}{\,\mathrm{m}E_h}

\begin{document}
\raggedbottom

\title{Classical Active-Space Hybrid Quantum Subspace Expansion (CASH-QSE): Quantum Corrections without Remeasuring the Classically Calculable Energy}

\author{Artur F. Izmaylov}
\email{artur.izmaylov@utoronto.ca}
\affiliation{Chemical Physics Theory Group, Department of Chemistry, University of Toronto, Toronto, Ontario M5S 3H6, Canada}
\affiliation{Department of Physical and Environmental Sciences, University of Toronto Scarborough, Toronto, Ontario M1C 1A4, Canada}

\date{\today}

\begin{abstract}
The wave function prepared by a variational quantum eigensolver (VQE) can contain a substantial classically tractable component, yet its energy contribution is sampled on the quantum device. Representing the wave function as a combination of a classical reference and separately prepared quantum states allows the reference energy to be evaluated without quantum sampling. However, near-linear dependence among these components can amplify measurement errors and destabilize the resulting energy estimate. We introduce Classical Active-Space Hybrid Quantum Subspace Expansion (CASH-QSE), which retains a complete-active-space self-consistent field (CASSCF) reference classically and uses occupation structure to construct quantum states exactly orthogonal to the reference and to one another. The energy follows from an ordinary Hermitian eigenvalue problem without overlap measurements, while the same occupation constraints simplify the measured operators. Using full configuration interaction to guide component selection, we test CASH-QSE along H$_2$O and N$_2$ bond-stretching coordinates spanning weakly to strongly correlated regimes. The benchmarks use STO--3G for both molecules and a restricted cc-pVDZ orbital space for H$_2$O. CASH-QSE reaches chemical accuracy while limiting the largest complete measurement circuits to a few hundred all-to-all logical controlled-NOT gates. In favorable cases, it also reduces the idealized final-energy sampling cost by several orders of magnitude relative to VQE with an adaptive derivative-assembled pseudo-Trotter ansatz (ADAPT-VQE).
\end{abstract}

\maketitle

\section{Introduction}
\label{sec:introduction}

Variational quantum eigensolvers (VQEs) face three coupled bottlenecks in
electronic-structure applications: preparing correlated states, optimizing
parametrized circuits, and collecting enough measurements to resolve energies
to chemical accuracy
\cite{Peruzzo2014,McClean2016Theory,Gonthier2022}.
On hardware without full error correction, state preparation is especially
limiting because errors accumulate with circuit depth and entangling-gate
count.  A useful variational strategy should therefore avoid placing the
entire correlation problem in one costly circuit while also controlling the
resulting optimization and measurement costs.

One way to reduce the burden on a single circuit is to distribute the wave
function among several independently prepared states.  Quantum-subspace
expansion (QSE), although introduced primarily for excited states and error
mitigation \cite{McClean2017QSE}, provides a natural framework for doing so:
\[
  \ket{\Psi}=\sum_{\mu=0}^{d-1}c_\mu\ket{\Phi_\mu}.
\]
Here \(d\) is the number of basis states \(\ket{\Phi_\mu}\), and \(c_\mu\) are their expansion coefficients. The global amplitudes follow from
\[
  \vct H\vct c=E\vct S\vct c,
  \qquad
  H_{\mu\nu}=\mel{\Phi_\mu}{\Hhat}{\Phi_\nu},
  \qquad
  S_{\mu\nu}=\braket{\Phi_\mu|\Phi_\nu}.
\]
Here \(\Hhat\) is the electronic Hamiltonian, and \(E\) is the Ritz energy.
This construction replaces part of the nonlinear optimization and deep state
preparation by a linear subspace problem.  Its price is the need to measure
additional pairwise matrix elements and, when the basis states are
nonorthogonal, the overlap matrix \(\vct S\).  Near-linear dependence can then
make the generalized eigenvalue problem sensitive to measurement errors
\cite{Motta2024Review,Kwao2026GEP}.

Nonorthogonal VQE, multireference quantum Krylov, circuit-generated, and
partitioned subspace methods explore different versions of this depth versus
subspace-size tradeoff
\cite{Huggins2020NOVQE,Stair2020MRSQK,Baek2023NOQE,
Cortes2022Krylov,Hirsbrunner2024CSVQE,OLeary2025PQSE}.
A related depth--measurement tradeoff occurs in low-depth unitary coupled
cluster through a Taylor expansion of small-amplitude factors \cite{Chen2022LowDepthUCC}.
Their common measurement burden suggests two related questions: can the basis
states be made orthogonal by construction, and can the same structure used to
enforce orthogonality also reduce the additional matrix-element measurements?

Quantum seniority-based subspace expansion (Q-SENSE) addresses both questions
\cite{Patel2026QSENSE}.  It uses orbital seniority—the pattern of singly
occupied orbitals—as a structural label even though the Hamiltonian does not
conserve seniority exactly.  States assigned to different seniority patterns
belong to disjoint subspaces and are therefore exactly orthogonal, so
\(\vct S=\vct I\).  The same labels identify Hamiltonian terms that cannot
connect a given pair of states, reducing the measurement cost created by the
larger subspace.

Orthogonality and structural measurement reduction do not, however, eliminate
another source of redundancy.  Both VQE and quantum-subspace methods may still
measure an energy contribution that is already calculable classically.
We consider
\[
  \ket{\Psi}
  =
  a\ket{C}
  +\sqrt{1-|a|^2}\ket{Q},
  \qquad
  \braket{C|Q}=0,
\]
where \(\ket{C}\) and \(\ket{Q}\) are normalized, \(\ket{C}\) is classically tractable, and \(\ket{Q}\) contains the remaining correlation. Its energy is
\[
  E
  =
  |a|^2E_C+(1-|a|^2)E_Q
  +2\,\operatorname{Re}\!\left[
  a^*\sqrt{1-|a|^2}\,H_{CQ}\right],
\]
where \(E_C=\mel{C}{\Hhat}{C}\), \(E_Q=\mel{Q}{\Hhat}{Q}\), and \(H_{CQ}=\mel{C}{\Hhat}{Q}\).
When \(|a|\simeq1\), the dominant term can be the already known classical
energy \(E_C\).  Preparing and measuring the complete correlated state on a
quantum device can therefore spend substantial resources remeasuring \(E_C\).

Classically boosted VQE (CB-VQE) avoids this redundancy by retaining an
explicit classically tractable state \cite{Radin2021CBVQE}.  In the
single-classical-state form relevant here,
\[
  \ket{C}=\ket{\mathrm{HF}},
  \qquad
  \ket{Q}=\hat U(\vct\theta)\ket{\mathrm{HF}},
  \qquad
  \braket{C|Q}\ne0,
\]
where \(\ket{\mathrm{HF}}\) is the Hartree--Fock determinant.  The classical
energy is evaluated without quantum sampling, while the remaining Hamiltonian
and overlap elements are measured and combined through a generalized
eigenvalue problem.  CB-VQE therefore removes the need to remeasure the
classical energy, but its nonorthogonal classical and quantum states retain the
overlap-measurement and conditioning problems.

The goal of preserving a classically tractable component while recovering
only its missing correlation is particularly relevant in multireference
chemistry.  Complete active-space self-consistent field (CASSCF) theory
\cite{Roos1980CASSCF} captures the dominant static correlation associated
with bond breaking and near-degeneracy, while correlation involving orbitals
outside the active space remains to be recovered.  Complete active-space
second-order perturbation theory (CASPT2) and $n$-electron valence-state
second-order perturbation theory (NEVPT2) add this external correlation
\cite{Andersson1990CASPT2,RoosAndersson1995LevelShift,Angeli2001NEVPT2}.
Quantum proposals have instead incorporated omitted orbitals through virtual
subspace expansion, localized active-space circuits, classical corrections,
or downfolded effective Hamiltonians
\cite{Takeshita2020VQSE,Urbanek2020VQSE,Otten2022LASUCC,
Matousek2024VQEAC,Singh2025DUCC}.

The unresolved design question is whether external correlation can be
recovered without remeasuring the classically calculable active-space block.
The correction should remain variational, and the quantum basis should remain
orthogonal so that no overlap matrix is required.  Classical Active-Space Hybrid Quantum Subspace
Expansion (\CASH) is designed to provide this combination. In its
single-classical-state form,
\begin{equation}
\begin{aligned}
  \ket{\Psi_{\rm CASH}}
  &=c_C\ket{C}+\sum_{a=1}^{N_Q}c_a\ket{q_a},\\
  \braket{C|q_a}&=0,
  \qquad
  \braket{q_a|q_b}=\delta_{ab}.
\end{aligned}
\label{eq:cash_ansatz_intro}
\end{equation}

Here \(N_Q\) is the number of retained quantum states.
The CASSCF component remains classically represented, and its internal
Hamiltonian block is evaluated classically.  The classical state may still
need to be prepared to measure its couplings to the quantum correction; what
CASH-QSE avoids is remeasuring its classically calculable internal block.  The
remaining correlation is divided among orthogonal, structurally constrained
quantum states that can be prepared and optimized separately.  Their global
combination follows from an ordinary Hermitian eigenvalue problem.  This
division exchanges one difficult state preparation for several simpler ones
and additional transition measurements, while the known structure of the
states helps limit the added measurement cost.

We test whether this classical/quantum partition can recover chemical accuracy
with favorable controlled-NOT (CNOT) gate counts and sampling costs for
H$_2$O and N$_2$ from
equilibrium to dissociation.  The H$_2$O benchmarks probe corrections outside
a deliberately restricted active space and within a larger cc-pVDZ orbital
window.  The N$_2$ benchmarks show how the resource balance changes with the
quality of the classical active-space reference.  These controlled, exact-information-assisted
statevector calculations establish attainable resource tradeoffs rather than
an end-to-end hardware implementation or a practical from-scratch
state-selection algorithm.

\section{Classical active space and exact quantum complement}
\label{sec:theory}

CASH-QSE separates the classical active space from its external complement
by requiring inactive orbitals to be doubly occupied and external virtual
orbitals to be empty.  Configurations that violate these conditions are
assigned to mutually orthogonal subspaces of the external complement.

\subsection{Electronic Hamiltonian and classical active space}

In an orthonormal spin-orbital basis, the electronic Hamiltonian is
\begin{equation}
  \Hhat
  =\sum_{ij} h_{ij}a_i^\dagger a_j
  +\frac{1}{4}\sum_{ijkl}g_{ijkl}
  a_i^\dagger a_j^\dagger a_l a_k,
  \label{eq:electronic_hamiltonian}
\end{equation}
Here \(a_i^\dagger\) and \(a_i\) are fermionic creation and annihilation
operators, \(h_{ij}\) are the one-electron integrals, and
\(g_{ijkl}=\langle ij\Vert kl\rangle\) are antisymmetrized two-electron
integrals \cite{Helgaker2000}.  In fermionic operator expressions, \(i,j,k,l\) denote spin orbitals and \(p,q,r,s\) spatial orbitals. A spin-orbital index is written as \(p\sigma\), with \(\sigma\in\{\alpha,\beta\}\), when the spin is explicit. Sector, state, and summation indices are specified separately.

Equation~\eqref{eq:electronic_hamiltonian} acts on the fermionic Fock space
\(\mathcal F\) generated by the chosen spin orbitals.  For the
nonrelativistic spin-independent Hamiltonians considered here, particle
numbers and molecular point-group symmetry define invariant blocks of
\(\mathcal F\).  We carry out the CASH-QSE construction in one such block,
\(\mathcal H_{\bm\xi}\subset\mathcal F\), where
\(\bm\xi=(N_\alpha,N_\beta,\Gamma_{\rm target})\) collects the fixed spin-resolved
particle numbers and target Abelian point-group irreducible representation.
This choice fixes spin projection but does not by itself fix total spin
\(S\). The final spin-adapted CASH components used in the benchmarks are
singlets; the general spin-orbital excitation pool need not conserve \(S\).

To identify the part of this symmetry sector that will remain classical,
we divide the spatial orbitals into inactive-core, active, and external virtual
sets,
\[
\begin{aligned}
  \mathcal C&=\{c_1,\ldots,c_{n_c}\},\qquad
  \mathcal A=\{a_1,\ldots,a_{n_o}\},\\
  \mathcal V&=\{v_1,\ldots,v_{n_v}\}.
\end{aligned}
\]
A complete active space (CAS) contains $n_e$ active electrons in $n_o$
spatial orbitals and is denoted CAS($n_e,n_o$), with \(n_o=|\mathcal A|\) and \(n_e=N_\alpha+N_\beta-2n_c\). Its subspace is identified
by the occupation operators
\[
\begin{aligned}
  \hat n_{p\sigma}&=a_{p\sigma}^\dagger a_{p\sigma},\\
  \hat D_p&=\hat n_{p\alpha}\hat n_{p\beta},\qquad
  \hat E_p=(1-\hat n_{p\alpha})(1-\hat n_{p\beta}).
\end{aligned}
\]
The operators \(\hat D_p\) and \(\hat E_p\) are Hermitian projectors that test
whether spatial orbital \(p\) is doubly occupied or empty, respectively.  The
CAS subspace \(\mathcal H_{\CAS}\subset\mathcal H_{\bm\xi}\) is defined by
\begin{equation}
\begin{aligned}
  \hat D_{c_u}\ket{\psi}&=\ket{\psi},
  && u=1,\ldots,n_c,\\
  \hat E_{v_a}\ket{\psi}&=\ket{\psi},
  && a=1,\ldots,n_v.
\end{aligned}
\label{eq:cas_occupation_conditions}
\end{equation}
Thus inactive orbitals are doubly occupied, external orbitals are empty, and
the active occupations remain unrestricted subject to the exact symmetries
in \(\bm\xi\).

The occupation conditions define the CAS subspace but not the particular
classical variational state within it.  We denote the normalized classical
reference by
\[
  \ket{C(\vct d)}
  =\sum_{I=1}^{d_{\CAS}}d_I\ket{D_I},
  \qquad
  \ket{C}\in\mathcal H_{\CAS},
  \qquad
  \braket{C|C}=1,
\]
where \(d_{\CAS}=\dim\mathcal H_{\CAS}\), \(\ket{D_I}\) are the CAS Slater determinants, and \(d_I\) are their coefficients.
For CASSCF, the coefficients \(\vct d\) and the orbitals are optimized
classically.  The HF limit is the one-dimensional special case
\(\mathcal H_{\CAS}=\Span\{\ket{\mathrm{HF}}\}\), with
\(\ket{C}=\ket{\mathrm{HF}}\).

\subsection{First-defect resolution of the external space}
\label{sec:first_defect}

CASH-QSE requires quantum states orthogonal to the classical CAS and to one
another.  Choosing a partition of the complement also determines which
fermionic generators can explore each sector without leaving it.  Resolving
all inactive-core and external-virtual spin-orbital occupation patterns makes
confinement straightforward, since fixed modes can be omitted from the
generators, but can produce up to \(2^{2(n_c+n_v)}\) sectors.  Conversely,
one or two complement sectors suffice if collective occupation conditions are
used.  These coarser partitions impose collective boundaries: a general
construction based on resolving aggregate occupation numbers introduces
controls whose fermionic rank can grow with system size.  The objective is
therefore to combine a small number of sectors with polynomial-size,
low-rank generator pools of sufficient expressivity.

Spatial-orbital occupation conditions provide a useful compromise between
sector count and generator complexity.  We obtain the partition directly
from the CAS occupation conditions in Eq.~\eqref{eq:cas_occupation_conditions}:
every occupation-basis configuration outside the CAS is assigned to the first
local CAS condition that it fails.  This ``first failed test'' defines its
\emph{first defect}.  Quantum basis states are then prepared within the
resulting nonempty subspaces, so states assigned to different first-defect
labels are orthogonal.

To make this assignment unique, we order the local occupation projectors as
\(\hat D_{c_1},\ldots,\hat D_{c_{n_c}},
\hat E_{v_1},\ldots,\hat E_{v_{n_v}}\) and write
\begin{equation}
  \hat X_j=
  \begin{cases}
    \hat D_{c_j}, & 1\le j\le n_c,\\
    \hat E_{v_k}, & j=n_c+k,\quad 1\le k\le n_v .
  \end{cases}
  \label{eq:ordered_constraints}
\end{equation}
For \(L=n_c+n_v\), the first-defect subspace \(\mathcal H_j\) consists of
states that satisfy every earlier CAS condition but fail the \(j\)th one:
\[
  \mathcal H_j
  =
  \left\{\ket{\psi}\in\mathcal H_{\bm\xi}:
  \begin{array}{l}
  \hat X_\ell\ket{\psi}=\ket{\psi}\quad(\ell<j),\\
  \hat X_j\ket{\psi}=0
  \end{array}\right\}.
\]

Every occupation-basis configuration in \(\mathcal H_{\bm\xi}\) either
passes all tests and belongs to \(\mathcal H_{\CAS}\), or has a unique first
failed test and belongs to one \(\mathcal H_j\).  Because the local occupation projectors
commute, the resulting subspaces form the exact orthogonal decomposition
\begin{equation}
  \mathcal H_{\bm\xi}
  =\mathcal H_{\CAS}\oplus\bigoplus_{j=1}^{L}\mathcal H_j.
  \label{eq:external_resolution}
\end{equation}
For example, if \(j<k\), every state in \(\mathcal H_j\) has eigenvalue zero
under \(\hat X_j\), whereas every state in \(\mathcal H_k\) has eigenvalue one
under the same Hermitian projector.  The CAS subspace has eigenvalue one under
every \(\hat X_j\).  The classical state and quantum states assigned to
different first-defect subspaces are therefore exactly orthogonal without
overlap measurements or an explicit orthogonalization step.

The ordering in Eq.~\eqref{eq:ordered_constraints} does not change the CAS
complement or the exactness of the direct-sum decomposition, but it does
redistribute configurations among the individual \(\mathcal H_j\) subspaces.
It is therefore a resource-design parameter: different fixed orderings can
change the state-preparation and measurement costs of the resulting
subspaces.  All calculations reported here use the ordering in
Eq.~\eqref{eq:ordered_constraints} without optimizing it.  No further local CAS
conditions are imposed after the first defect, so \(\mathcal H_j\) can contain configurations
of different excitation ranks and should not be interpreted as an
excitation-rank class.

States within \(\mathcal H_{\CAS}\) that are orthogonal to the current
\(\ket{C}\) are not introduced as separate quantum basis states.  Their
variational freedom is retained by optimizing the classical CAS coefficients
or by retaining multiple classical Ritz states within the CAS.
CASH-QSE therefore treats
internal active-space correlation through the classical component and
reserves the quantum basis for its external complement.

\section{Low-CNOT quantum states in orthogonal sectors}
\label{sec:quantum_states}

The first-defect decomposition solves the orthogonality problem, but it does
not by itself provide short circuits for representing the quantum complement.
We therefore construct each retained quantum state inside one of the
first-defect subspaces, using only generators that preserve its defining
first-defect occupation conditions.  If a first-defect subspace is too broad
to represent compactly, it can be split into smaller orthogonal subspaces
before state preparation.  We call this further partitioning
\emph{structural refinement}; its purpose is to reduce the preparation cost
of each state at the expense of a larger subspace basis.

\subsection{First-defect seeds and sector-preserving circuits}
\label{sec:implementable_circuits}

For every nonempty first-defect subspace \(\mathcal H_j\), we choose a normalized
determinant or compact spin-adapted configuration state function (CSF)
\(\ket{s_j}\in\mathcal H_j\).  Thus,
\(\hat X_\ell\ket{s_j}=\ket{s_j}\) for \(\ell<j\) and
\(\hat X_j\ket{s_j}=0\).  

Starting from \(\ket{s_j}\), a quantum state is generated as
\begin{align*}
  \ket{q_j(\vct\theta_j)}
  &=\hat U_j(\vct\theta_j)\ket{s_j},\\
  \hat U_j(\vct\theta_j)
  &=\prod_{m=1}^{K_j}
  \exp\!\left(\theta_{jm}\hat A_{jm}\right),
  \qquad
  \hat A_{jm}^\dagger=-\hat A_{jm}.
\end{align*}
Here \(K_j\) is the number of unitary factors and \(\theta_{jm}\) are real circuit parameters. The parent pool \(\mathcal P\) contains generalized anti-Hermitian fermionic
single and double excitation generators,
\begin{align*}
  \hat A_{ij}^{(1)}&=a_i^\dagger a_j-a_j^\dagger a_i,\\
  \hat A_{ijkl}^{(2)}&=a_i^\dagger a_j^\dagger a_l a_k
  -a_k^\dagger a_l^\dagger a_j a_i,
\end{align*}
restricted to the common particle-number, spin-projection, and spatial
symmetries.  The first-defect pool
\(\mathcal P_j\subseteq\mathcal P\) is the subset whose generators preserve
\(\mathcal H_j\).  A sufficient algebraic test is
\begin{align*}
  [\hat A,\hat X_\ell]&=0,\qquad \ell<j,\\
  \hat X_j\hat A(1-\hat X_j)
  &=(1-\hat X_j)\hat A\hat X_j=0,
\end{align*}
together with preservation of the common exact symmetries.  Thus every
accepted generator preserves its first-defect subspace.
The elementwise restriction \(\mathcal P_j\subseteq\mathcal P\) should be
distinguished from allowing leakage-free linear combinations in
\(\Span\mathcal P\).  The latter can retain transformations lost by
individual screening.  Appendix~\ref{app:sector_control} compares alternative
occupation resolutions and analyzes the controllability of such an augmented
pool.

\subsection{Structural refinement by seniority and fixed occupations}
\label{sec:seniority_refinement}

A broad first-defect subspace can be split by assigning fixed eigenvalues
to additional commuting occupation and seniority operators.  These eigenvalues
label the smaller orthogonal subspaces.  For spatial orbital \(p\), the local
seniority operator is
\[
  \hat\Omega_p
  =\hat n_{p\alpha}+\hat n_{p\beta}
  -2\hat n_{p\alpha}\hat n_{p\beta},
  \qquad
  \hat\Omega_p^2=\hat\Omega_p .
\]
It has eigenvalue one for a singly occupied orbital and zero for an empty or
doubly occupied orbital.  The smaller subspaces can also be distinguished by
fixed eigenvalues of the spatial occupation operator
\[
  \hat N_p=\hat n_{p\alpha}+\hat n_{p\beta},
  \qquad N_p\in\{0,1,2\}.
\]

A label \(\lambda\) records the chosen seniority and occupation eigenvalues
within \(\mathcal H_j\).  We denote their
joint eigenspace by
\(\mathcal H_{j\lambda}\subseteq\mathcal H_j\).  For constrained orbitals,
\begin{align*}
  \hat\Omega_p\ket{\psi}
  &=\omega_p^{(\lambda)}\ket{\psi},
  \qquad \omega_p^{(\lambda)}\in\{0,1\},\\
  \hat N_r\ket{\psi}
  &=\eta_r^{(\lambda)}\ket{\psi},
  \qquad \eta_r^{(\lambda)}\in\{0,1,2\}.
\end{align*}
The restricted generators preserve the assigned eigenspaces; sufficient
conditions are \([\hat A,\hat N_p]=0\) for each constrained occupation and
\([\hat A,\hat\Omega_p]=0\) for each constrained seniority.  The same eigenvalue
argument used for the first-defect partition makes distinct refined subspaces
orthogonal.  For each refined label, \(\ket{s_{j\lambda}}\in\mathcal H_{j\lambda}\) is a normalized seed and \(\hat U_{j\lambda}\) its sector-preserving preparation unitary. Since \(\hat U_{j\lambda}\) preserves \(\mathcal H_{j\lambda}\),
independent optimization gives
\[
  \langle s_{j\lambda}|\hat U_{j\lambda}^{\dagger}
  \hat U_{j\eta}|s_{j\eta}\rangle=0,\qquad\lambda\ne\eta.
\]
The calculations reported here retain at most one independently optimized
quantum state for each final structural label.  The formal construction does
not prescribe the order in which individual \(\hat\Omega_p\) or
\(\hat N_p\) constraints are introduced. Refinement is guided by the
chosen preparation-cost criterion. 
This refinement need not be implemented by projecting a previously optimized
broad state.  A broad state may instead be replaced by separately prepared states in the
smaller invariant subspaces.

\paragraph{Exact partition and variational representation.}
Because the first-defect decomposition and its structural refinements are
exact orthogonal partitions, any real state in the target symmetry space can
be written as
\[
  \ket{\Psi}=\ket{\Psi_{\rm CAS}}
  +\sum_{j,\lambda}\ket{\Psi_{j\lambda}},
  \qquad \ket{\Psi_{j\lambda}}\in\mathcal H_{j\lambda},
\]
Here \(\ket{\Psi_{\rm CAS}}\in\mathcal H_{\CAS}\), and the sum includes all nonempty refined subspaces. Within each
subspace, quantum states are prepared using the single- and double-excitation
generators that preserve the imposed symmetries. We do not establish that
the restricted generator pools can represent every state in these subspaces.
Their expressivity is supported empirically by the high-fidelity preparation
of the retained full configuration interaction (FCI) components in the molecular
benchmarks described in Sec.~\ref{sec:results} and the Supplementary
Information (SI).

\section{Hybrid Ritz problem and self-consistent optimization}
\label{sec:ritz}

The global amplitudes follow from an ordinary Hermitian Ritz problem,
coupled to optimization of the quantum-state parameters and the classical
component.

\subsection{Contracted orthogonal problem}

The set \(\Lambda_j\) contains the retained structured labels in first-defect
subspace \(j\).  In the unrefined construction \(\Lambda_j\) contains one
label and \(\ket{q_{j\lambda}}\equiv\ket{q_j}\).  The CASH-QSE basis is
\begin{align*}
  \mathcal B_{\mathrm{CASH}}
  &=\{\ket{C}\}\cup
  \{\ket{q_{j\lambda}}:\ j=1,\ldots,L,\ 
  \lambda\in\Lambda_j\},\\
  d_{\mathrm{CASH}}&=1+\sum_{j=1}^{L}|\Lambda_j|,
\end{align*}
where \(d_{\mathrm{CASH}}\) is the total number of basis states.  Every
\(\ket{q_{j\lambda}}\) is normalized and belongs to
\(\mathcal H_{j\lambda}\).  Their assigned subspaces give
\[
  \braket{C|q_{j\lambda}}=0,\qquad
  \braket{q_{j\lambda}|q_{k\eta}}=\delta_{jk}\delta_{\lambda\eta}.
\]
Thus the overlap matrix is the identity.

The global amplitudes are obtained
from the ordinary Hermitian matrix
\[
  \vct H_{\mathrm{hyb}}
  =\begin{pmatrix}
    E_C & \vct h^\dagger\\
    \vct h & \vct H_Q
  \end{pmatrix},
\]
where, with the composite quantum index \(a=(j,\lambda)\),
\[
\begin{aligned}
  E_C&=\mel{C}{\Hhat}{C},\qquad
  h_a=\mel{q_a}{\Hhat}{C},\\
  (H_Q)_{ab}&=\mel{q_a}{\Hhat}{q_b}.
\end{aligned}
\]
Solving
\(\vct H_{\mathrm{hyb}}\vct c=E\vct c\) yields the coefficients
\(c_C\) and \(c_a\) in the CASH-QSE wave function of
Eq.~\eqref{eq:cash_ansatz_intro}.  Thus the global mixing of the classical and
quantum components is linear even when the individual quantum states are
optimized nonlinearly.

The same construction permits \(N_{\rm cl}\) mutually orthonormal classical
states \(\ket{C_k}\in\mathcal H_{\CAS}\), with \(\ket{C_1}=\ket{C}\) when
\(N_{\rm cl}=1\). The Ritz dimension then becomes
\(d_{\rm CASH}=N_{\rm cl}+N_Q\), with \(N_Q=\sum_j|\Lambda_j|\), and the
scalar \(E_C\) is replaced by the classically evaluated block
\((H_{\rm cl})_{kl}=\mel{C_k}{\Hhat}{C_l}\). Each classical coefficient
is an independent Ritz variable. This differs from decomposing one
contracted reference solely to measure its transitions. The cc-pVDZ
benchmarks use this multiple-classical-state construction.

\subsection{Adaptive construction and amplitude optimization}
\label{sec:adaptive_construction}

The Ritz problem provides the global objective for selecting a generator
and the quantum state to which its exponential is applied.  For a current quantum state
\(\ket{q_a}\in\mathcal H_{j\lambda}\),
\(\mathcal P_a\equiv\mathcal P_{j\lambda}\) denotes the corresponding
sector-preserving generator pool obtained from the parent pool in
Sec.~\ref{sec:implementable_circuits}, including any additional constraints
introduced in Sec.~\ref{sec:seniority_refinement}.  For a candidate generator
\(\hat A\in\mathcal P_a\), appending \(\exp(t\hat A)\) gives the screening gradient
\[
  g_{\hat A}^{(a)}
  =
  \left.
  \frac{d}{dt}\lambda_{\min}[\vct H_{\mathrm{hyb}}(t)]
  \right|_{t=0}.
\]
Here \(t\) is a real trial parameter and \(\lambda_{\min}\) denotes the smallest eigenvalue. For a nondegenerate lowest Ritz root, the Hellmann--Feynman expression gives
\[
  g_{\hat A}^{(a)}
  =|c_a|^2\mel{q_a}{[\Hhat,\hat A]}{q_a}
  +2\operatorname{Re}\sum_{\nu\ne a}c_\nu^*c_a
  \mel{\phi_\nu}{\Hhat\hat A}{q_a},
\]
Here \(\ket{\phi_\nu}\) denotes either a classical Ritz state \(\ket{C_k}\) or a quantum state \(\ket{q_b}\); the sum excludes the varied state \(\ket{q_a}\).  Screening therefore measures the effect of a candidate generator on
the full hybrid energy rather than on the isolated diagonal energy of one
state.

Following the adaptive operator-selection philosophy of the adaptive
derivative-assembled pseudo-Trotter ansatz VQE (ADAPT-VQE)
\cite{Grimsley2019ADAPT}, the global iteration is:
\begin{enumerate}
  \setlength{\itemsep}{0.35em}
  \item The current hybrid Ritz matrix is constructed and diagonalized to obtain
  \(E\) and \(\vct c\).
  \item The gradient \(g_{\hat A}^{(a)}\) is evaluated for every eligible
  \(\hat A\in\mathcal P_a\) and every retained \(\ket{q_a}\).
  \item The pair \((\ket{q_a},\hat A)\) with the largest
  \(|g_{\hat A}^{(a)}|\) is selected, with exact ties resolved by a fixed pool order.
  The corresponding unitary factor is appended to that state's preparation circuit.
  \item All accepted circuit amplitudes are reoptimized by minimizing the lowest
  Ritz root.
  \item The iteration continues until the chosen energy/gradient criterion is
  satisfied. If a state's preparation cost exceeds a chosen resource cap, the
  affected broad state is replaced using the structural refinement of
  Sec.~\ref{sec:seniority_refinement}.
\end{enumerate}
The numerical convergence tolerances and preparation-cost cap are
implementation parameters.  The resource benchmarks use FCI-assisted state
selection rather than executing this adaptive loop from scratch. 

The joint amplitude optimization accounts for the coupling of all retained
states through the Ritz matrix.  Its energy derivatives follow from the
Hellmann--Feynman derivative of the hybrid matrix.

\subsection{CASSCF coefficient and orbital macrocycles}
\label{sec:casscf_macrocycle}

The contracted CASSCF state can relax in response to the quantum correction.
At fixed orbitals and quantum states, its coefficients are optimized by
retaining the full classical CAS block in the Ritz problem.

The determinants \(\{\ket{D_I}\}\) span the complete active space, and
\(\ket{q_a}\) runs over the retained quantum states.  The matrix elements are
\[
  (H_{\rm CAS})_{IJ}=\mel{D_I}{\Hhat}{D_J},
  \qquad
  V_{Ia}=\mel{D_I}{\Hhat}{q_a}.
\]
For fixed orbitals and quantum states, the CAS coefficients are updated by
solving
\begin{equation}
  \begin{pmatrix}
    \vct H_{\rm CAS} & \vct V\\
    \vct V^\dagger & \vct H_Q
  \end{pmatrix}
  \begin{pmatrix}\vct d\\\vct c_Q\end{pmatrix}
  =E\begin{pmatrix}\vct d\\\vct c_Q\end{pmatrix}.
  \label{eq:uncontracted_problem}
\end{equation}
Here \(\vct d\) and \(\vct c_Q\) are, respectively, the CAS and quantum blocks
of the normalized total hybrid eigenvector.  The CAS block therefore need not
itself have unit norm. For \(\|\vct d\|>0\), before it is used as the contracted
classical state in the next Ritz iteration, it is normalized:
\begin{equation}
  \widetilde{\vct d}=\frac{\vct d}{\|\vct d\|},
  \qquad
  \ket{\widetilde C}=\sum_I\widetilde d_I\ket{D_I}.
  \label{eq:normalized_cas_update}
\end{equation}
The norm \(\|\vct d\|\) is then carried by the corresponding global Ritz
coefficient.

The formal method can also relax the orbitals through derivatives of the
hybrid Ritz energy, as in related orbital-optimized VQE approaches
\cite{Fitzpatrick2024SCF}.  The orbital parameterization and full
self-consistent macrocycle are given in Appendix~\ref{app:orbital_macrocycle}.
Orbital relaxation is not used in the reported resource benchmarks.

\section{Matrix elements and structure-based measurement reduction}
\label{sec:measurement}

Measurement of the enlarged Ritz matrix can be costly
\cite{Choi2023ExcitedMeasurement,Patel2025MeasurementReview}; we combine
diagonal and transition estimators with structural operator reductions to
lower its sampling cost.

\subsection{Matrix-element measurements and classical-branch implementations}
\label{sec:extended_swap}

The purely classical Hamiltonian block is evaluated classically, so the
quantum device is needed only for quantum-state diagonal elements and
classical--quantum or quantum--quantum transitions.  For an off-diagonal
matrix element between normalized basis states \(\ket{\phi_\mu}\) and
\(\ket{\phi_\nu}\), we use the interference state
\begin{equation}
  \ket{\Phi_{\mu\nu}}
  =\frac{1}{\sqrt2}
  \left(\ket{0}\ket{\phi_\mu}+\ket{1}\ket{\phi_\nu}\right).
  \label{eq:extended_swap_state}
\end{equation}
For any system operator \(\hat O_{\mu\nu}\),
\begin{equation}
  \mel{\phi_\mu}{\hat O_{\mu\nu}}{\phi_\nu}
  =\mel{\Phi_{\mu\nu}}
  {(X+\mathrm iY)\otimes\hat O_{\mu\nu}}
  {\Phi_{\mu\nu}}.
  \label{eq:extended_swap_general}
\end{equation}
Here \(X\) and \(Y\) are the ancilla Pauli operators.
The measured observables must be Hermitian even when
\(\hat O_{\mu\nu}\) is not. With
\[
 \hat O_R=\frac{\hat O_{\mu\nu}+\hat O_{\mu\nu}^\dagger}{2},\qquad
 \hat O_I=\frac{\hat O_{\mu\nu}-\hat O_{\mu\nu}^\dagger}{2\mathrm i},
\]
the observables \(X\otimes\hat O_R-Y\otimes\hat O_I\) and
\(Y\otimes\hat O_R+X\otimes\hat O_I\) yield the real and imaginary
transition components, respectively. For Hermitian \(\hat O_{\mu\nu}\),
these reduce to \(X\otimes\hat O_{\mu\nu}\) and
\(Y\otimes\hat O_{\mu\nu}\). This is the extended swap-test estimator
used in Q-SENSE \cite{Patel2026QSENSE,Parrish2019QFD}.

\begin{figure*}[t]
\centering
\begin{tikzpicture}[x=1.35cm,y=1.05cm,font=\small,line width=0.8pt]
  \draw (0,0) -- (7.4,0);
  \draw (0,-1.1) -- (7.4,-1.1);
  \draw (0,-2.2) -- (7.4,-2.2);
  \node[left] at (0,0) {$a:\ |0\rangle$};
  \node[left] at (0,-1.1) {$A:\ |0^n\rangle$};
  \node[left] at (0,-2.2) {$B:\ |0^n\rangle$};

  \node[draw,minimum width=0.55cm,minimum height=0.5cm,fill=white] at (0.75,0) {$H$};

  \draw (1.65,0) -- (1.65,-1.1);
  \draw[fill=white] (1.65,0) circle (0.085);
  \node[draw,minimum width=0.78cm,minimum height=0.48cm,fill=white] at (1.65,-1.1) {$\hat C_\mu$};

  \draw (2.55,0) -- (2.55,-2.2);
  \fill (2.55,0) circle (0.085);
  \node[draw,minimum width=0.78cm,minimum height=0.48cm,fill=white] at (2.55,-2.2) {$\hat C_\nu$};

  \node[draw,minimum width=1.1cm,minimum height=0.48cm,fill=white] at (3.55,0) {$P_a(-\delta)$};
  \node[draw,minimum width=0.78cm,minimum height=0.48cm,fill=white] at (3.55,-1.1) {$\hat U_\mu$};
  \node[draw,minimum width=0.78cm,minimum height=0.48cm,fill=white] at (3.55,-2.2) {$\hat U_\nu$};

  \draw[densely dashed] (4.25,0.35) -- (4.25,-2.55);
  \node[above,align=center] at (4.25,0.42)
    {\scriptsize branch-selected\\[-1pt]\scriptsize preparations};

  \draw (5.15,0) -- (5.15,-2.2);
  \fill (5.15,0) circle (0.085);
  \draw (5.02,-0.97) -- (5.28,-1.23);
  \draw (5.02,-1.23) -- (5.28,-0.97);
  \draw (5.02,-2.07) -- (5.28,-2.33);
  \draw (5.02,-2.33) -- (5.28,-2.07);
  \node[above,align=center] at (5.15,0.25)
    {\scriptsize CSWAP$^{\otimes n}$};

  \draw[densely dashed] (5.85,0.35) -- (5.85,-2.55);
  \node[above] at (6.55,0.26) {\scriptsize \(X/Y\) readout};
  \node[above] at (6.55,-0.85) {\scriptsize \(P_\ell\) readout};
  \node[above] at (6.55,-1.95) {\scriptsize discarded};
  \node[right] at (7.4,-2.2) {$|0^n\rangle$};

  \node[anchor=west,align=left] at (0,-3.0)
  {\scriptsize \(\circ\): control on \(a=0\);\quad
   \(\bullet\): control on \(a=1\).\quad
   \(\hat C_{\mu/\nu}\): controlled seed/occupation preparation;\quad
   \(\hat U_{\mu/\nu}\): number-conserving ansatz.};
\end{tikzpicture}
\caption{Two-register extended-swap circuit for off-diagonal CASH-QSE matrix
elements. Here \(H\) denotes the Hadamard gate, and \(P_\ell\) a system Pauli word. The ancilla selects the seed/occupation preparation on register
\(A\) or \(B\).  The ansatz circuits need not be controlled because the
logical idle state \(\ket{0^n}\) is verified to be invariant under the executed
tapered ansatz circuits up to phase.  The ancilla gate
\(P_a(-\delta)\) compensates any difference between the two
idle-state phases and is omitted when that difference is zero modulo
\(2\pi\).  The CSWAP network maps both
branches onto register \(A\), where ancilla \(X/Y\) and system-Pauli
measurements recover the real and imaginary transition components.}
\label{fig:extended_swap_circuit}
\end{figure*}
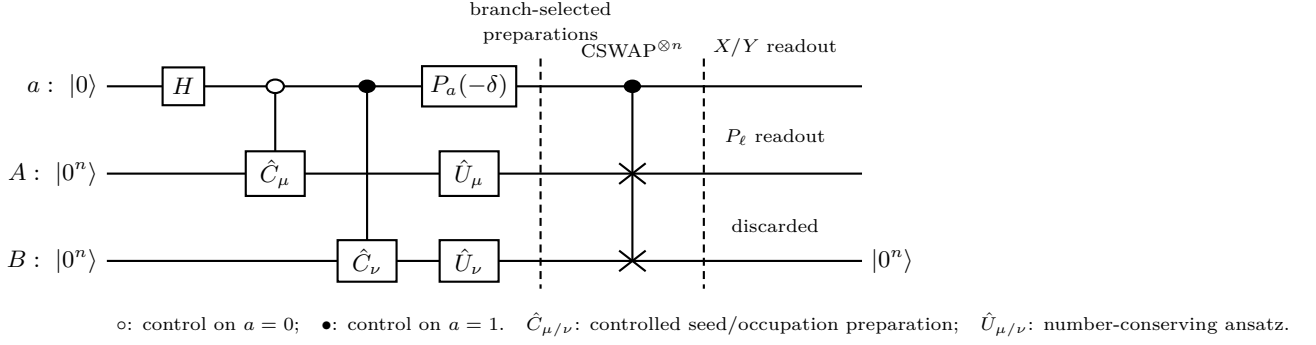

To avoid controlling an entire correlated-state preparation, the same
interference state can be realized with two system registers and a
controlled-SWAP (CSWAP) network, as shown in
Fig.~\ref{fig:extended_swap_circuit}.  Here \(\hat C_\mu\) and
\(\hat C_\nu\) denote the branch-controlled seed/occupation preparations, and
\(\hat U_\mu,\hat U_\nu\) denote the subsequent number-conserving ansatz
circuits.  In the untapered fermionic representation these ansatz circuits
leave the physical vacuum invariant.  After tapering to a fixed symmetry
sector, however, the logical idle state need not represent the physical
vacuum.  The uncontrolled-ansatz construction instead requires a common
logical idle state \(\ket{\eta}\) satisfying
\[
  \hat U_\rho^{(\mathrm{tap})}\ket{\eta}
  =e^{\mathrm i\varphi_\rho}\ket{\eta},
  \qquad \rho\in\{\mu,\nu\}.
\]
For all reported tapered transition circuits, idle-state invariance is
verified with \(\ket{\eta}=\ket{0^n}\), where \(n\) is the number of qubits in each tapered system register.  Before correcting any relative
phase, the controlled swap disentangles this idle register and leaves the
ancilla and measurement register in
\[
\ket{\Phi_{\mu\nu}(\delta)}=
\frac{|0\rangle|\phi_\mu\rangle+
e^{\mathrm i\delta}|1\rangle|\phi_\nu\rangle}{\sqrt2},
\]
where \(\delta=\varphi_\mu-\varphi_\nu\).  The ancilla phase gate
\(P_a(-\delta)=\operatorname{diag}(1,e^{-\mathrm i\delta})\) restores
\(\ket{\Phi_{\mu\nu}}\) for the estimator in
Eq.~\eqref{eq:extended_swap_general}.  It can be placed before or after the
controlled swap, adds no CNOTs, and is omitted when \(\delta=0\) modulo
\(2\pi\).

A complete off-diagonal measurement circuit therefore contains both state
preparations, the branch-controlled seed operations, the controlled-SWAP
network, and the basis-change circuit required to measure a mutually commuting
group, or fragment, of Pauli terms.  This complete circuit, rather than a
state-preparation circuit alone, is the relevant two-qubit-gate-count measure
for a transition matrix element.

A multiconfigurational classical state introduces a second practical issue:
its coherent preparation may itself be expensive even though its amplitudes
are known classically.  If a compact coherent preparation is available, we write
\begin{equation}
  \ket{C}
  =\hat U_{\rm CAS}\ket{\mathrm{HF}}
  =\sum_I d_I\ket{D_I}.
  \label{eq:coherent_cas_branch}
\end{equation}
Alternatively, the same normalized Ritz vector can be decomposed classically
into mutually orthogonal structured components,
\[
  \ket{C}=\sum_{k=1}^{N_C^{\rm comp}}\alpha_k\ket{C_k},
  \qquad
  \braket{C_k|C_l}=\delta_{kl},
\]
where \(N_C^{\rm comp}\) is the number of measurement components and \(\alpha_k=\braket{C_k|C}\) are their amplitudes. Each transition is reconstructed from
\begin{equation}
  \mel{C}{\Hhat}{q_a}
  =\sum_k\alpha_k^*\mel{C_k}{\Hhat}{q_a}.
  \label{eq:component_resolved_cas_transition}
\end{equation}
In Eq.~\eqref{eq:component_resolved_cas_transition}, the component states
implement measurements of one contracted classical Ritz state. When they
are instead retained as independent classical Ritz states, their coefficients
are optimized directly as described in Sec.~\ref{sec:ritz}.

\subsection{Removing nonconnecting Hamiltonian terms}
\label{sec:connected_operators}

Structural constraints identify Hamiltonian monomials that cannot connect a
given pair of basis states.  We remove these monomials to obtain a
pair-specific connected operator. For a contracted reference, with
\(\ket{\phi_0}=\ket{C}\) and
\(\ket{\phi_a}=\ket{q_a}\) for quantum indices \(a\),
\(\mathcal S_\mu\) denotes the support used for this test.  For the coherent
classical state, \(\mathcal I_C=\{I:d_I\ne0\}\) and
\(\mathcal S_0=\{D_I:I\in\mathcal I_C\}\).
For a quantum state, \(\mathcal S_a\) is the certified determinant/CSF support
implied by its structural label, rather than its unknown amplitude support.

Writing \(\Hhat=\sum_{\alpha\in\mathcal T_1\cup\mathcal T_2}
h_\alpha\hat\tau_\alpha\), where \(\mathcal T_1\) and \(\mathcal T_2\)
index one- and two-electron monomials, we define
\[
\begin{aligned}
\mathcal A_{\mu\nu}=\{\alpha:\;&\exists D_\mu\in\mathcal S_\mu,
\ D_\nu\in\mathcal S_\nu\ \text{with}\\
&\mel{D_\mu}{\hat\tau_\alpha}{D_\nu}\ne0\},\\
\hat T_{\mu\nu}^{\rm conn}&=\sum_{\alpha\in\mathcal A_{\mu\nu}}
h_\alpha\hat\tau_\alpha.
\end{aligned}
\]
For classical--quantum and quantum--quantum pairs, including diagonals,
\[
\mel{\phi_\mu}{\Hhat}{\phi_\nu}
=\mel{\phi_\mu}{\hat T_{\mu\nu}^{\rm conn}}{\phi_\nu}.
\]
Component-resolved classical transitions and independent classical Ritz
states use the same test separately for each \(\ket{C_k}\).  Additional occupation or seniority constraints shrink
the certified support and can only remove further monomials.

This is an algebraic support test.  For \(N_{\rm so}\) spin orbitals,
\(N_C=|\mathcal I_C|\), and \(N_Q=\sum_j|\Lambda_j|\), its classical cost is
bounded by \(\order(N_C N_{\rm so}^4)\) for one classical--quantum pair,
\(\order(N_C N_Q N_{\rm so}^4)\) for all such pairs, and
\(\order(N_Q^2N_{\rm so}^4)\) for all quantum--quantum pairs. For
multiple classical Ritz states, \(N_C\) is replaced by the sum of their
determinant-support sizes. Broad
first-defect masks can be cached before applying finer constraints.
Exact linear relations among the surviving monomials permit further
reductions that preserve the required matrix elements.

\subsection{Exact transition-null operators from local constraints}
\label{sec:killers}

We call an operator whose matrix element between the selected bra and ket
vanishes exactly a transition-null, or ``killer,'' operator.  To be useful for
measurement reduction, such an operator must also remain cheap enough to
measure.  Expectation-preserving changes of measured operator
decompositions have previously been used to reduce variance, for example
through ghost Pauli products \cite{Choi2022Ghost}.  CASH-QSE instead uses
occupation and seniority operators minus their fixed eigenvalues times the
identity.  These residual operators annihilate one member of the chosen
bra--ket pair.  Products of many residuals would have high fermionic body rank,
so the candidate killer operators use only residual operators of low
fermionic body rank.

For each \(\ket{\phi_\mu}\), the fixed eigenvalues are
\(\nu_{p\sigma}^{(\mu)}\in\{0,1\}\),
\(\eta_p^{(\mu)}\in\{0,1,2\}\), and
\(\omega_p^{(\mu)}\in\{0,1\}\) for the previously defined
operators \(\hat n_{p\sigma}\), \(\hat N_p\), and \(\hat\Omega_p\), respectively.
We denote the fixed spin-orbital and seniority index sets by
\(\mathcal F_\mu\) and \(\mathcal W_\mu\).  Only constraints that are exact
for the prepared state are used to construct candidate killer operators.

To obtain a flexible two-body null space from a fixed one-body occupation,
we multiply the residual occupation operator by a symmetry-allowed one-electron operator
\[
  \hat O_{p\sigma}^{X}(\vct o)
  =o_{p\sigma,0}^{X}I
  +\sum_{rs,\tau}o_{p\sigma,rs\tau}^{X}
  a_{r\tau}^\dagger a_{s\tau},
  \qquad X\in\{L,R\}.
\]
For a matrix element between \(\ket{\phi_\mu}\) and
\(\ket{\phi_\nu}\), the left- and right-annihilating contributions are
\begin{align*}
  \hat K_{\mu\nu}^{L}
  &=\sum_{(p,\sigma)\in\mathcal F_\mu}
  \big(\hat n_{p\sigma}-\nu_{p\sigma}^{(\mu)}I\big)
  \hat O_{p\sigma}^{L},\\
  \hat K_{\mu\nu}^{R}
  &=\sum_{(p,\sigma)\in\mathcal F_\nu}
  \hat O_{p\sigma}^{R}
  \big(\hat n_{p\sigma}-\nu_{p\sigma}^{(\nu)}I\big).
\end{align*}
The same construction applies to \(\hat N_p-\eta_p^{(\mu)}I\) on the left and \(\hat N_p-\eta_p^{(\nu)}I\) on the right for fixed spatial occupations.  Since \(\hat\Omega_p\)
is already a two-electron operator, seniority residual operators are multiplied
only by scalars,
\[
\begin{aligned}
  \hat K_{\mu\nu}^{\Omega,L}
  &=\sum_{p\in\mathcal W_\mu}b_p^L
  \big(\hat\Omega_p-\omega_p^{(\mu)}I\big),\\
  \hat K_{\mu\nu}^{\Omega,R}
  &=\sum_{p\in\mathcal W_\nu}b_p^R
  \big(\hat\Omega_p-\omega_p^{(\nu)}I\big).
\end{aligned}
\]
The residual is always placed next to the state it annihilates, so each term
has zero target matrix element.

The coefficients \(o_{p\sigma,0}^X\), \(o_{p\sigma,rs\tau}^X\), and \(b_p^X\), with \(X=L,R\), are the free killer coefficients, collected in \(\vct o\). Collecting all contributions into \(\hat K_{\mu\nu}(\vct o)\) gives the exact invariance
\begin{equation}
  \mel{\phi_\mu}{\Hhat}{\phi_\nu}
  =
  \mel{\phi_\mu}
  {\hat T_{\mu\nu}^{\mathrm{conn}}
   -\hat K_{\mu\nu}(\vct o)}
  {\phi_\nu}.
  \label{eq:killer_invariance}
\end{equation}
For diagonal matrix elements the corresponding Hermitian killer operators
are formed with anticommutators of the occupation residuals and Hermitian one-electron
multipliers.  For off-diagonal elements, orthogonality also makes the identity
an exact transition-null operator.
After the nonidentity coefficients are chosen, the shifted residual is
\(\hat R_{\mu\nu}=\hat R_{\mu\nu}^{(0)}+o_{\mu\nu}^{I,*}I\).
The superscript \(*\) on \(o_{\mu\nu}^{I,*}\) denotes the optimal value. The shift minimizing the variance of the unfragmented real-transition estimator is
\begin{equation}
  o_{\mu\nu}^{I,*}
  =-\frac12\operatorname{Re}\!\left[
  \mel{\phi_\mu}{\hat R_{\mu\nu}^{(0)}}{\phi_\mu}
  +\mel{\phi_\nu}{\hat R_{\mu\nu}^{(0)}}{\phi_\nu}
  \right],
  \qquad \mu\ne\nu,
  \label{eq:constant_shift}
\end{equation}
with
\(\hat R_{\mu\nu}^{(0)}
=\hat T_{\mu\nu}^{\mathrm{conn}}-\hat K_{\mu\nu}\).
The diagonal expectations in this expression depend on the prepared states;
the benchmarks evaluate them from the retained statevectors.

\subsection{Structure-informed one-norm killer optimization}
\label{sec:one_norm}

The linear combination of candidate killer operators introduces free
coefficients. The nonidentity coefficients are chosen without using the
many-body amplitudes of the target state.  We map the connected operator and every elementary killer operator to a
common Pauli basis. For off-diagonal elements, the expansions and optimization below use the Hermitian extended-swap representation of Sec.~\ref{sec:extended_swap}:
\[
  \hat T_{\mu\nu}^{\mathrm{conn}}
  =\sum_\ell t_\ell^{\mu\nu}P_\ell,
  \qquad
  \hat K_r^{\mu\nu}
  =\sum_\ell B_{\ell r}^{\mu\nu}P_\ell.
\]
Here \(P_\ell\) are Pauli words, \(r\) labels elementary killer operators, and \(t_\ell^{\mu\nu}\) and \(B_{\ell r}^{\mu\nu}\) are their real expansion coefficients. We define
\[
  \hat R_{\mu\nu}(\vct o)
  =\hat T_{\mu\nu}^{\mathrm{conn}}
  -\sum_r o_r\hat K_r^{\mu\nu}
  =\sum_\ell r_\ell^{\mu\nu}(\vct o)P_\ell .
\]
Equation~\eqref{eq:killer_invariance} guarantees that every choice of
\(\vct o\) preserves the Ritz matrix element.

We choose the real coefficients \(o_r\) by minimizing the unweighted Pauli one-norm:
\begin{align*}
  \min_{\vct o,\vct u}\quad&\sum_\ell u_\ell,\\
  \text{subject to}\quad&
  -u_\ell\le
  t_\ell^{\mu\nu}-\sum_rB_{\ell r}^{\mu\nu}o_r
  \le u_\ell,\qquad u_\ell\ge0.
\end{align*}
Here the auxiliary variables \(u_\ell\) bound the coefficient magnitudes.
For an off-diagonal element the optimization is applied to the Hermitian
extended-swap dilation and followed by the analytic identity shift in
Eq.~\eqref{eq:constant_shift}. For each off-diagonal element, the benchmarks
compare the connected operator, its identity-shifted form, and the shifted
one-norm solution. Diagonal elements omit the redundant identity-only
candidate. Each candidate is grouped separately, and the one with the
smallest sum of fragment standard deviations is retained. Thus exact
variances enter candidate selection as well as final shot allocation.
Covariance fitting is excluded from the final protocol.

The one-norm objective is motivated by the pre-grouping bound
\[
  M_{\mu\nu}\lesssim
  \frac{\left(\sum_\ell|r_\ell^{\mu\nu}|\right)^2}
       {\epsilon_{\mu\nu}^2},
\]
where \(M_{\mu\nu}\) is the shot count and \(\epsilon_{\mu\nu}\) the target standard error of the matrix element.
The nonidentity one-norm coefficients depend only on the known structural
constraints and operator coefficients. Learning the diagonal expectations
for the identity shift and the variances for candidate selection and shot
allocation incurs additional measurements in a hardware implementation.

\subsection{Finite-sampling error of the CASH-QSE Ritz energy}
\label{sec:sampling_error}

Shot allocation accounts for both fragment variances and the sensitivity of
the final Ritz energy \cite{Crawford2021SI,Patel2025MeasurementReview}.
For independent estimators with variances \(\sigma_i^2/M_i\), the
minimum-variance allocation for a weighted sum with weights \(w_i\) is
\(M_i\propto|w_i|\sigma_i\).  We apply this rule to fragments, classical
components, and Ritz matrix elements.

The benchmarks use real orbitals, Hamiltonians, and circuit parameters, so
the Ritz matrix and its eigenvectors are real.  Complex problems require
separate allocation and propagation of real and imaginary transition
estimators.  For the final selected Hermitian measured operator (including any identity shift),
\(\hat R'_{\mu\nu}=\sum_g\hat F_{\mu\nu}^{(g)}\), including the ancilla
Pauli for an off-diagonal element, \(\sigma_{\mu\nu}^{(g)}\) denotes the
single-shot standard deviation of the measured fragment \(\hat F_{\mu\nu}^{(g)}\). Hats on estimated quantities, such as \(\widehat H_{\mu\nu}\), denote statistical estimators.

The set \(\mathcal Q\) contains the independent upper-triangular Ritz elements that
require quantum measurement. The entire classically known block is excluded:
\(H_{00}=E_C\) for a contracted reference, or \(\vct H_{\rm cl}\) for
multiple classical Ritz states. Pairs with identically zero connected
operators are also excluded.  Continuous optimal allocation of
\(M_{\mu\nu}=\sum_gM_{\mu\nu}^{(g)}\) shots among the fragments of one matrix
element gives
\begin{equation}
  \operatorname{Var}[\widehat H_{\mu\nu}]
  =\frac{\sigma_{\mu\nu}^2}{M_{\mu\nu}},
  \qquad
  \sigma_{\mu\nu}=\sum_g\sigma_{\mu\nu}^{(g)}.
  \label{eq:matrix_element_variance}
\end{equation}

For component-resolved classical transitions, the same rule gives
\(\sigma_{ka}=\sum_g\sigma_{ka}^{(g)}\) for each component and, with
independent component estimates and \(M_{Ca}=\sum_kM_{ka}\),
\begin{equation}
\begin{aligned}
  M_{ka}
  &=M_{Ca}\,
  \frac{|\alpha_k|\sigma_{ka}}
       {\sum_l|\alpha_l|\sigma_{la}},\\
  \operatorname{Var}[\widehat H_{Ca}]
  &=\frac{(\sigma_{Ca}^{\rm eff})^2}{M_{Ca}},
  \qquad
  \sigma_{Ca}^{\rm eff}
  =\sum_k|\alpha_k|\sigma_{ka}.
\end{aligned}
  \label{eq:component_resolved_allocation}
\end{equation}
The effective coefficient \(\sigma_{Ca}^{\rm eff}\) replaces
\(\sigma_{Ca}\) in the Ritz-level allocation.  For normalized ground-state
Ritz coefficients \(\vct c\) and a nondegenerate lowest root, first-order eigenvalue perturbation theory
assigns sensitivities \(c_\mu^2\) to diagonal elements and
\(2c_\mu c_\nu\) to independent upper-triangular off-diagonal elements.
Applying the allocation rule at total budget
\(M=\sum_{(\mu,\nu)\in\mathcal Q}M_{\mu\nu}\) gives the squared sampling standard error \(\epsilon_{\mathrm{CASH}}^2\), to this order,
\begin{equation}
  \begin{aligned}
  \epsilon_{\mathrm{CASH}}^2
  &=\frac{\mathcal C_{\mathrm{CASH}}}{M},\\
  \mathcal C_{\mathrm{CASH}}
  &=\left[
  \sum_{\mu:(\mu,\mu)\in\mathcal Q}c_\mu^2\sigma_{\mu\mu}
  +2\!\sum_{\substack{\mu<\nu\\(\mu,\nu)\in\mathcal Q}}
  |c_\mu c_\nu|\sigma_{\mu\nu}
  \right]^2 .
  \end{aligned}
  \label{eq:cash_sampling_metric}
\end{equation}
The optimal budgets are proportional to
\(c_\mu^2\sigma_{\mu\mu}\) for diagonal elements and
\(2|c_\mu c_\nu|\sigma_{\mu\nu}\) for off-diagonal elements.

These are ideal continuous allocations with known fragment variances.
Group floors \(M_{\mu\nu}^{(g)}\ge m_{\min}\) account for a minimum number
of shots per group.
Pilot measurements, used to estimate fragment variances and allocate the
final sampling budget, are excluded.  For a target final-energy sampling standard error \(\epsilon_{\mathrm{meas}}\), the VQE comparator obeys
\[
\mathcal C_{\mathrm{VQE}}=\left(\sum_g\sigma_g^{\mathrm{VQE}}\right)^2,
\qquad
M_{\mathrm{VQE}}=\mathcal C_{\mathrm{VQE}}/\epsilon_{\mathrm{meas}}^2,
\]
while \(M_{\mathrm{CASH}}=\mathcal C_{\mathrm{CASH}}/
\epsilon_{\mathrm{meas}}^2\).  This metric covers only the converged
final-energy estimate.  Intermediate Ritz matrices, generator screening,
state optimization, and orbital updates require additional measurements.

\section{Optional exact-symmetry tapering of the complete protocol}
\label{sec:exact_tapering}

Exact-symmetry tapering reduces the register while preserving the common
sector used for the states, Hamiltonian, generators, and measured operators.

The operators \(\{\hat G_a\}_{a=1}^{r}\) are independent commuting Pauli symmetries of
the mapped Hamiltonian,
\[
  [\hat G_a,\Hhat]=0,\qquad
  [\hat G_a,\hat G_b]=0,
\]
with target eigenvalues \(g_a=\pm1\).  Every classical and quantum basis state
used in one Ritz problem is restricted to this common sector,
\begin{equation}
  \hat G_a\ket{\phi_\mu}=g_a\ket{\phi_\mu},
  \qquad
  \ket{\phi_\mu}\in\{\ket{C_k}\}_{k=1}^{N_{\rm cl}}\cup\{\ket{q_b}\}.
  \label{eq:common_exact_sector}
\end{equation}
The construction in Sec.~II of Ref.~\cite{Setia2020Tapering} yields a Clifford
unitary \(\hat C_{\rm tap}\) that maps the symmetry generators to single-qubit
Paulis whose eigenvalues can be fixed and whose qubits can then be removed.  If \(\hat V_{\bm g}\) is the isometry from the tapered register
into the full register with the transformed symmetry qubits fixed to their
target eigenstates, the physical-sector embedding is
\(\hat C_{\rm tap}^{\dagger}\hat V_{\bm g}\). A symmetry-preserving operator
\(\hat O\) is represented by
\[
  \hat O^{(\bm g)}
  =
  \hat V_{\bm g}^{\dagger}
  \hat C_{\rm tap}\hat O\hat C_{\rm tap}^{\dagger}
  \hat V_{\bm g}.
\]
Because the same isometry is applied to every Ritz basis state, inner
products and Hamiltonian matrix elements are preserved.

Pair-specific connected operators and killer operators need not be written
term by term in a symmetry-commuting Pauli basis before tapering.  If
\(\hat O=\sum_\ell o_\ell P_\ell\), we retain only the Pauli words that commute
with every exact symmetry,
\begin{equation}
  \hat O_{\parallel}
  =
  \sum_{\ell:\,[P_\ell,\hat G_a]=0\ \forall a}
  o_\ell P_\ell.
  \label{eq:exact_sector_pauli_filter}
\end{equation}
Any Pauli word that anticommutes with at least one \(\hat G_a\) has zero
matrix element between states satisfying
Eq.~\eqref{eq:common_exact_sector}; hence
\begin{equation}
  \bra{\phi_\mu}\hat O\ket{\phi_\nu}
  =
  \bra{\phi_\mu}\hat O_{\parallel}\ket{\phi_\nu}.
  \label{eq:exact_sector_filter_invariance}
\end{equation}
This is a classical deletion of symmetry-forbidden Pauli words, not a
state-space projection performed on the quantum device.

The same mapping, symmetry filter, and tapering transformation are applied
throughout the protocol.  For the two-register transition circuit, the
logical idle state must remain a common eigenstate of the tapered ansatz
circuits, with any relative idle-state phase compensated as in
Sec.~\ref{sec:extended_swap}.  Physical-vacuum invariance before tapering is
insufficient.  Tapering is optional; omitting it retains the equivalent
untapered representation.

\section{Computational details}
\label{sec:computational}

\subsection{Molecular models and classical references}
\label{sec:molecular_spaces}

Classical calculations provide the reference states and orbital bases, while
FCI supplies benchmark energies and information for selecting quantum
components.  Restricted Hartree--Fock (RHF), state-specific CASSCF, and FCI
calculations use the PySCF electronic-structure package \cite{Sun2018PySCF},
symmetry-adapted molecular orbitals, fixed spin populations, and the totally
symmetric irreducible representation of the molecular point group.
CASH-QSE with an HF reference (HF-CASH) uses RHF orbitals; CASH-QSE with a
CASSCF reference (CAS-CASH) uses the corresponding stored CASSCF orbitals.
The reference orbitals remain fixed during the resource benchmarks.

\paragraph{H$_2$O/STO--3G.}
Both O--H bonds are stretched symmetrically through
\(R_{\mathrm{OH}}=0.96,\ 1.75,\ 3.0\)~\AA{} at a fixed H--O--H angle of
\(104.5^\circ\).  The minimal STO--3G basis contains seven spatial orbitals
and ten electrons, with point group \(C_{2v}\).  The classical reference is
either the closed-shell HF determinant or CASSCF(2,2).  The latter retains
the \(A_1\) bond-breaking pair and leaves the complementary \(B_2\) pair
outside the active space, providing a controlled test of quantum corrections
to an incomplete bond-breaking reference.  A CAS(4,4) diagnostic includes
both pairs.

\paragraph{N$_2$/STO--3G.}
Calculations at \(R_{NN}=1.1,\ 1.8,\ 3.0\)~\AA{} use ten spatial orbitals
and 14 electrons in the Abelian \(D_{2h}\) group.  The classical references
are the closed-shell HF determinant, CASSCF(4,4), and CASSCF(6,6).
CAS(4,4) contains the two \(\pi/\pi^*\) bond-breaking pairs; CAS(6,6) also
contains the \(\sigma/\sigma^*\) pair and therefore all three valence
bond-breaking channels.  The CAS(2,2) overlap diagnostic contains only the
\(\sigma/\sigma^*\) pair.  These choices test how the chemistry captured
classically affects the quantum correction.

\paragraph{H$_2$O/cc-pVDZ extension.}
The larger calculation uses a restricted space from the correlation-consistent
polarized valence double-zeta (cc-pVDZ) basis at the same water geometries.
The \(1a_1\) core orbital is frozen, leaving eight electrons in 14 spatial
orbitals: two retained inactive orbitals, the four CAS(4,4) orbitals containing
both bond-breaking pairs, and eight external virtual orbitals.
The STO--3G active pair and the retained cc-pVDZ orbitals are tracked between
geometries by maximum overlap among orbitals belonging to the same
irreducible representation.  Explicit orbital lists and ordering conventions
are given in Appendix~\ref{app:computational_specs}.

\subsection{Mapping and symmetry reduction}

Fermionic Hamiltonians, excitation generators, connected operators, and
killer operators are constructed before Jordan--Wigner mapping,
exact-symmetry filtering, and tapering.  Spin and point-group occupation
parities reduce the registers from 14 to 10 qubits for H$_2$O/STO--3G,
20 to 15 for N$_2$/STO--3G, and 28 to 24 for H$_2$O/cc-pVDZ.
CASH-QSE and its ADAPT-VQE comparator use the same exact symmetry sector and
tapering transformation.  All resource quantities are recomputed in the
final tapered representation.  The parity operators, target eigenvalues,
and orbital-rotation restrictions are specified in
Appendix~\ref{app:computational_specs}.

\subsection{State construction and comparison protocol}

The benchmarks use FCI information to select structured components, rather
than executing the formal adaptive loop from scratch; state-selection costs
are excluded from the reported resources.
For H$_2$O/STO--3G, one adaptive state is initially constructed in each
nonempty first-defect subspace from the symmetry-allowed determinant of lowest
diagonal Hamiltonian energy.  States exceeding 100 tapered preparation CNOTs
are replaced by states with resolved seniorities and, when necessary, fixed
occupations.  Component selection continues until the final Ritz energy
satisfies \(|E-E_{\mathrm{FCI}}|\le1.6\mEh\). Here \(E_{\mathrm{FCI}}\) is the benchmark FCI energy and \(E_h\) the hartree.
The initial basis before refinement is diagnostic and need not satisfy
chemical accuracy. Initial and final spaces need not be nested because
refinement can replace states, impose singlet adaptation, select components,
and contract contributions.

N$_2$/STO--3G and H$_2$O/cc-pVDZ use direct construction of the final
structured basis under the same 100-CNOT cap.  Fixed paired occupations
provide an additional subdivision in cc-pVDZ.  No universal order for adding
occupation and seniority constraints is assumed.  The seed and generator
choices for the different seniority classes are given in
Appendix~\ref{app:computational_specs}.

The preparation cap also determines how classical--quantum transitions are
measured.  The H$_2$O CAS(2,2) reference is prepared coherently; the N$_2$
CAS(4,4) and CAS(6,6) references are prepared through orthogonal components
with fixed open-shell and paired-occupation patterns.  Their transitions
reconstruct the contracted classical-state matrix element through
Eq.~\eqref{eq:component_resolved_cas_transition}.  The N$_2$ CAS(4,4)
benchmark additionally relaxes the CAS coefficients through
Eq.~\eqref{eq:uncontracted_problem} at fixed orbitals.  The full
orbital/circuit macrocycle is not used for the reported resource ratios.

For H$_2$O/cc-pVDZ, an uncontracted CAS-coefficient update at fixed CASSCF
orbitals is followed by decomposition of the updated CAS vector into eleven
orthogonal occupation/seniority components. These remain independent
classical Ritz states, giving \(d_{\rm CASH}=11+N_Q\). Their full
classical--classical Hamiltonian block is evaluated classically; only their
couplings to the quantum correction require quantum measurement.

One fermionic ADAPT-VQE comparator in the HF orbital basis is used per
geometry for all CASH-QSE references.  For H$_2$O/cc-pVDZ, it uses the
tracked RHF orbital basis and the same pool of generalized spin-orbital
single- and double-excitation generators at every geometry.  The comparator
is the shortest chemically accurate ADAPT-VQE circuit found under that pool,
tapering map, and synthesis convention, including FCI-targeted searches.

\subsection{Circuit and sampling resources}
\label{sec:measurement_protocol}

CNOT counts are obtained from explicit circuit constructions with all-to-all
logical connectivity.  A complete off-diagonal measurement circuit includes
both preparations, controlled seed synthesis, the controlled-SWAP network,
and the basis change for its fully commuting Sorted Insertion (FC-SI)
fragment \cite{Yen2020FC,Crawford2021SI}.  Classical-reference preparations
are counted whether coherent or component resolved; each nonzero component
transition is measured independently.

For every quantum-measured diagonal and transition element, the connected
operator and the full sets of left- and right-state killer operators are
constructed from fixed occupations and local seniority constraints.
Diagonal elements use Hermitian combinations of the killer operators.
The structure-informed one-norm optimization follows the common symmetry
filter and tapering. Candidate selection and the identity shift follow
Sec.~\ref{sec:one_norm}.
Exact statevector fragment variances determine both the retained candidate
and the idealized continuous allocation at the sampling-uncertainty target
\(\epsilon_{\mathrm{meas}}=1.6\mEh\), including component-resolved
classical transitions.

The CASH-QSE total sums all FC-SI groups of all nonzero quantum-measured Ritz
elements; the exactly known classical block and algebraically zero connected
operators contribute no shots.  ADAPT-VQE uses all FC-SI groups of its
Hamiltonian expectation value under the same variance convention.
For H$_2$O/cc-pVDZ at 1.75~\AA{}, the variances use the accepted 459-factor ADAPT-VQE state, matching the reported circuit. At 3.00~\AA{}, the ADAPT-VQE
sampling estimate uses exact window FCI as a proxy, while its circuit cost
refers to the finite 68-factor state; the SI documents its fidelity and
energy error.
The reported sampling totals exclude pilot shots, integer group minima,
learning the identity shifts and fragment variances, gradient screening, amplitude and orbital
optimization measurements, noise, and hardware routing.  The H$_2$O/cc-pVDZ
check imposing at least one shot per measurement group is given in
Appendix~\ref{app:discrete_allocations}.

Statevector checks verify symmetry preservation, agreement of tapered and
untapered matrix elements and energies, and the transition-circuit
estimators.  The implementation checks and inventory of numerical records
are given in Appendix~\ref{app:computational_specs}.

\section{\texorpdfstring{Results}{Results}}
\label{sec:results}

Bond stretching changes both the adequacy of the classical reference and the
correlation left for quantum treatment.  The benchmarks ask how these changes
affect preparation and measurement costs.  The H$_2$O calculations first test
an active space that omits part of the bond-breaking correlation, then a
balanced active space with a larger external orbital space.  The N$_2$
calculations test the importance of including both the \(\sigma\) and \(\pi\)
bonds in the classical description.
The comparisons use the common resource conventions in Computational details.

Table~\ref{tab:classical_fci_overlaps} shows how well each classical reference
represents the FCI wave function, through
\(F_{\rm ref}=|\langle\Psi_{\rm ref}|\Psi_{\rm FCI}\rangle|^2\).
Each FCI vector is expressed in the orbital basis of its corresponding reference.

\begin{table*}[t]
\centering
\caption{Squared classical-reference/FCI overlaps \(F_{\rm ref}\) for the
H$_2$O/STO--3G and N$_2$/STO--3G benchmarks.  Here \(R=R_{\rm OH}\) for H$_2$O
and \(R=R_{NN}\) for N$_2$.  Primary CASH-QSE references are HF and CAS(2,2)
for H$_2$O and HF, CAS(4,4), and CAS(6,6) for N$_2$; the remaining entries are
active-space diagnostics.}
\label{tab:classical_fci_overlaps}
\begin{ruledtabular}
\begin{tabular}{lccccc}
System & $R$ (\AA) & HF & CAS(2,2) & CAS(4,4) & CAS(6,6) \\
\hline
H$_2$O & 0.96 & 0.973 & 0.979 & 0.998 & -- \\
H$_2$O & 1.75 & 0.645 & 0.734 & 0.999 & -- \\
H$_2$O & 3.0 & $8.7\times10^{-5}$ & $8.0\times10^{-5}$ & $>0.999$ & -- \\
N$_2$ & 1.1 & 0.917 & 0.918 & 0.974 & 0.989 \\
N$_2$ & 1.8 & 0.426 & 0.444 & 0.837 & 0.995 \\
N$_2$ & 3.0 & 0.0716 & 0.129 & 0.371 & $>0.999$ \\
\end{tabular}
\end{ruledtabular}
\end{table*}

\subsection{\texorpdfstring{H$_2$O: accurate bond dissociation from simple quantum components}{H2O: accurate bond dissociation from simple quantum components}}

Symmetric O--H stretching exposes the limitation of CAS(2,2), which retains
only one of the two symmetry-adapted bond-breaking pairs.  At dissociation,
both HF and CAS(2,2) have negligible FCI overlap
(Table~\ref{tab:classical_fci_overlaps}).  The FCI weight in the
optimized CAS(2,2) subspace is itself only \(1.7\times10^{-4}\), so changing
its coefficients cannot repair the missing bond-breaking description.
The quantum complement must therefore supply essential configurations.

Across the three geometries, CASH-QSE nevertheless reaches chemical accuracy
with 7--11 basis states and at most 96 CNOTs per quantum preparation
(Table~\ref{tab:h2o_accuracy_circuits}).  Resolving the correction by
occupation and spin allows a strongly correlated total wave function to be
assembled from individually simple components.

\begin{table*}[t]
\caption{H$_2$O/STO--3G CNOT resources before and after
seniority/fixed-occupation refinement.  The initial one-state-per-first-defect basis is diagnostic and
need not satisfy chemical accuracy; \(d_{\rm CASH}\) is the final chemically accurate Ritz
dimension.  ``Prep'' denotes the largest state-preparation cost and ``full''
the largest complete final-energy measurement circuit.  The final column is
ADAPT-VQE full/refined CASH-QSE full.  Geometry and ADAPT baseline entries
span the rows at each geometry.}
\label{tab:h2o_accuracy_circuits}
\begin{center}
\resizebox{\textwidth}{!}{%
\begin{tabular}{lccccccccc}
\hline\hline
Reference & $R_{\rm OH}$ (\AA) & $d_{\rm CASH}$ & Broad prep & Refined prep & ADAPT prep &
Broad full & Refined full & ADAPT full & ADAPT/CASH\\
\hline
HF & \multirow{2}{*}{0.96} & 9 & 1009 & 83 & \multirow{2}{*}{1424} & 1674 & 240 & \multirow{2}{*}{1434} & 6.0\\
CAS(2,2) &  & 7 & 477 & 96 &  & 667 & 250 &  & 5.7\\
\midrule
HF & \multirow{2}{*}{1.75} & 11 & 769 & 96 & \multirow{2}{*}{1400} & 1211 & 251 & \multirow{2}{*}{1412} & 5.6\\
CAS(2,2) &  & 9 & 637 & 96 &  & 987 & 266 &  & 5.3\\
\midrule
HF & \multirow{2}{*}{3.0} & 7 & 152 & 48 & \multirow{2}{*}{2440} & 378 & 174 & \multirow{2}{*}{2452} & 14\\
CAS(2,2) &  & 8 & 256 & 64 &  & 587 & 212 &  & 12\\
\hline\hline
\end{tabular}}
\end{center}
\end{table*}

The largest circuit advantage over ADAPT-VQE occurs at dissociation, despite
the poor classical-reference overlap.  ADAPT-VQE builds the correlated wave
function coherently from a single determinant, whereas CASH-QSE prepares its
components separately and determines their relative amplitudes classically.
A large quantum contribution therefore need not require a long circuit for
each component.  Refinement enforces the preparation budget; at dissociation,
the components also require substantially fewer CNOTs than the cap permits,
while the ADAPT-VQE circuit becomes longer.

The savings survive the complete transition-measurement overhead.  Relative
to the initial broad representation, refinement distributes the preparation
task over a larger basis.  The broad and refined bases need not be nested or
equally accurate, so their circuit costs compare different variational
representations.  The resulting increase in measured matrix elements can
offset the savings from short preparations.

\subsection{Reference quality and occupation constraints jointly determine sampling cost}

The H$_2$O sampling costs vary nonmonotonically as the classical reference
deteriorates (Table~\ref{tab:h2o_measurement_results}).  Near
equilibrium, the classical state carries most of the wave function, reducing
the sensitivity of the final energy to quantum measurements.  At intermediate
stretching, more weight enters the quantum correction, while killers provide
little additional reduction of the connected-only sampling cost.

\begin{table*}[t]
\caption{Idealized H$_2$O/STO--3G final-energy sampling costs at
\(\epsilon_{\mathrm{meas}}=1.6\mEh\).  Here \(|c_C|^2\) is the optimized
classical-state weight, and CASH-QSE costs include occupation and seniority
killers.  The final column is ADAPT-VQE/CASH-QSE.  Geometry and ADAPT baseline
entries span the rows at each geometry.}
\label{tab:h2o_measurement_results}
\begin{ruledtabular}
\begin{tabular}{lcccccc}
Reference & $R_{\rm OH}$ (\AA) & $|c_C|^2$ & FC-SI groups &
CASH shots & ADAPT shots & ADAPT/CASH\\
\hline
HF & \multirow{2}{*}{0.96} & 0.974 & 134 & $5.6\times10^{3}$ & \multirow{2}{*}{$2.9\times10^6$} & 520\\
CAS(2,2) &  & 0.980 & 82 & $9.0\times10^{3}$ &  & 330\\
\midrule
HF & \multirow{2}{*}{1.75} & 0.652 & 215 & $2.6\times10^{5}$ & \multirow{2}{*}{$3.0\times10^6$} & 12\\
CAS(2,2) &  & 0.735 & 149 & $2.1\times10^{5}$ &  & 14\\
\midrule
HF & \multirow{2}{*}{3.0} & $5.7\times10^{-5}$ & 66 & $3.2\times10^{2}$ & \multirow{2}{*}{$1.2\times10^5$} & 370\\
CAS(2,2) &  & $5.6\times10^{-5}$ & 91 & $3.2\times10^{2}$ &  & 360\\
\end{tabular}
\end{ruledtabular}
\end{table*}

At dissociation, the classical weight is smaller still, yet the sampling cost
drops sharply.  Killers reduce the HF-CASH connected-only cost by about
\(175\times\), with a similar reduction for CAS(2,2).  Most of the CASH-QSE
wave-function weight lies in a quantum component with four singly occupied
orbitals forming two nearly triplet electron pairs coupled to an overall
singlet.  Occupation killers allow the diagonal and spin-flip parts of the
leading exchange interactions to be measured together.  Their fluctuations
then largely cancel, reducing the sampling cost despite the negligible
classical-reference weight. The SI reports the pair-spin correlations and
a fixed-state control in which restoring the original fragment membership
removes the dominant diagonal benefit.  The dissociation advantage thus comes mainly
from exploitable structure in the quantum component, whereas the equilibrium
advantage benefits from a large classical contribution.  Reference overlap
alone cannot predict both regimes.  The smallest shot totals remain
continuous estimates; the effect of a minimum allocation per group is tested
in the larger-orbital-space benchmark.

\subsection{\texorpdfstring{H$_2$O/cc-pVDZ: correlation beyond a bond-breaking active space}{H2O/cc-pVDZ: correlation beyond a bond-breaking active space}}
\label{sec:h2o14_ccpvdz_results}

CAS(4,4) includes both O--H bond-breaking pairs and captures nearly the entire
STO--3G FCI wave function (Table~\ref{tab:classical_fci_overlaps}).  Enlarging
the external orbital space now tests recovery of correlation beyond this
balanced valence description.  At dissociation, enlarging the cc-pVDZ window
from eight to fourteen orbitals increases the energy missing from CAS(4,4)
from 0.16 to 22~\(\mEh\) (Table~\ref{tab:h2o14_classical}).  The classical
reference still retains 97--99\% squared overlap with FCI across the
14-orbital geometries.  Capturing the principal bond-breaking configurations
therefore does not remove the need to recover energetically important
correlation involving the external orbitals.

\begin{table*}[t]
\caption{Effect of enlarging the retained H$_2$O/cc-pVDZ orbital window at
fixed CAS(4,4).  Here 8o and 14o are the total retained spatial-orbital
windows, while the active space is unchanged.  \(\Delta E_{\rm CAS}=E_{\rm CASSCF}-E_{\rm FCI}\)
measures the correlation left outside the classical reference within the same
window, and \(F_{\rm CAS}=|\langle\Psi_{\rm CASSCF}|\Psi_{\rm FCI}\rangle|^2\).}
\label{tab:h2o14_classical}
\begin{ruledtabular}
\begin{tabular}{ccccc}
$R_{\rm OH}$ (\AA)&$\Delta E_{\rm CAS}$, 8o ($\mathrm{m}E_h$)&$F_{\mathrm{CAS}}$, 8o&$\Delta E_{\rm CAS}$, 14o ($\mathrm{m}E_h$)&$F_{\mathrm{CAS}}$, 14o\\
\hline
0.96&5.2&0.997&79&0.973\\
1.75&2.7&0.998&39&0.981\\
3.0&0.16&$>0.999$&22&0.991
\end{tabular}
\end{ruledtabular}
\end{table*}

CASH-QSE recovers this correlation to chemical accuracy using hundreds of basis
states at the shorter geometries, yet each quantum preparation needs at most
61 CNOTs and complete measurement circuits stay below 400 CNOTs
(Table~\ref{tab:h2o14_resources}).  The external correlation is distributed
among many simple components, with the associated cost appearing in the
classical eigenvalue problem and the additional quantum matrix elements.

\begin{table*}[t]
\caption{CASH-QSE and ADAPT-VQE resources for the 14-orbital H$_2$O/cc-pVDZ
window. CASH-QSE uses fixed CASSCF orbitals and a CAS(4,4) partition with relaxed classical coefficients; ADAPT-VQE uses the tracked RHF orbitals. All reported states satisfy chemical accuracy.
Here \(d_{\rm CASH}=11+N_Q\) includes eleven independent classical
components, ``full'' denotes the largest
complete final-energy measurement circuit, and the final column is the
ADAPT-VQE/CASH-QSE shot ratio.  The maximum individual quantum-state
preparation cost is 61 CNOTs at every geometry. The 3.0~\AA{} ADAPT shot
estimate uses the FCI proxy specified in Computational details.}
\label{tab:h2o14_resources}
\begin{center}
\resizebox{0.96\textwidth}{!}{%
\begin{tabular}{cccccc}
\hline\hline
$R_{\rm OH}$ (\AA)&$d_{\rm CASH}$&CASH full CNOT&ADAPT-VQE full CNOT&CASH shots&ADAPT/CASH shot ratio\\
\hline
0.96&282&383&92228&$2.1\times10^4$&$1.4\times10^3$\\
1.75&229&375&81556&$1.2\times10^4$&$1.7\times10^3$\\
3.0&25&350&8660&$5.1\times10^2$&$8.2\times10^3$\\
\hline\hline
\end{tabular}}
\end{center}
\end{table*}

At dissociation, chemical accuracy requires only 25 basis states: eleven
classical components and fourteen external quantum states. This smaller
basis accompanies a decrease in the CASSCF--FCI energy gap from
\(79\mEh\) near equilibrium to \(22\mEh\)
(Table~\ref{tab:h2o14_classical}). The active natural occupations approach
one, consistent with an open-shell bond-breaking description within
CAS(4,4). Because the active space includes the principal bond-breaking
configurations, stronger correlation in the full wave function need not
increase the quantum correction. At a fixed absolute accuracy, a larger
fraction of the smaller reference energy deficit can remain unrecovered,
helping explain why fewer components can suffice. These observations do
not establish whether the external correction also becomes more
concentrated among the selected components. Natural occupations and
solver diagnostics are provided in the Supplementary Information.

The sampling advantage survives a minimum of one shot per commuting group
on both methods: ADAPT-VQE/CASH-QSE ratios remain approximately 560--4,400
(Appendix~\ref{app:discrete_allocations}).  These estimates still assume known
variances and exclude pilot measurements.

\subsection{\texorpdfstring{N$_2$: treating the triple bond in the active space}{N2: treating the triple bond in the active space}}
\label{sec:n2_results}

For N$_2$, CAS(4,4) describes the two \(\pi\) bond-breaking pairs but leaves
the \(\sigma/\sigma^*\) pair outside the classical reference.  CAS(6,6)
includes all three.  The overlap improvements in
Table~\ref{tab:classical_fci_overlaps} reflect the recovery of these
bond-breaking configurations.  Comparing the circuit and sampling costs
then tests which resources benefit from the improved reference.  CAS(6,6)
still needs a quantum correction at the two shorter bond lengths, but already
reaches chemical accuracy at dissociation, the ``classical stop'' in
Tables~\ref{tab:n2_circuit_results} and~\ref{tab:n2_measurement_results}.

\begin{table*}[t]
\caption{N$_2$/STO--3G circuit resources at chemical accuracy.  ``Prep''
denotes state preparation and ``full'' the largest complete final-energy
measurement circuit.  Every executed CASH-QSE preparation requires at most
99 CNOTs.  The final column is ADAPT-VQE full/CASH-QSE full; ``classical
stop'' means that the classical reference already meets the accuracy target.
Geometry and the common ADAPT baseline span the rows at each geometry.}
\label{tab:n2_circuit_results}
\begin{ruledtabular}
\begin{tabular}{lccccc}
Reference & $R_{NN}$ (\AA) & CASH full & ADAPT prep & ADAPT full & ADAPT/CASH\\
\hline
HF & \multirow{3}{*}{1.1} & 343 & \multirow{3}{*}{9192} & \multirow{3}{*}{9222} & 27\\
CAS(4,4) &  & 334 &  &  & 28\\
CAS(6,6) &  & 327 &  &  & 28\\
\midrule
HF & \multirow{3}{*}{1.8} & 342 & \multirow{3}{*}{21548} & \multirow{3}{*}{21577} & 63\\
CAS(4,4) &  & 333 &  &  & 65\\
CAS(6,6) &  & 328 &  &  & 66\\
\midrule
HF & \multirow{3}{*}{3.0} & 338 & \multirow{3}{*}{6040} & \multirow{3}{*}{6066} & 18\\
CAS(4,4) &  & 328 &  &  & 19\\
CAS(6,6) &  & -- &  &  & classical stop\\
\end{tabular}
\end{ruledtabular}
\end{table*}

All cases requiring quantum corrections reach chemical accuracy under the
same 100-CNOT preparation cap.  The narrow range of complete circuit costs in
Table~\ref{tab:n2_circuit_results} therefore reflects the common component
construction as well as the chemistry of the reference.  At 1.8~\AA{},
replacing HF by CAS(6,6) changes the largest complete CASH-QSE circuit only
from 342 to 328 CNOTs, while the sampling advantage over ADAPT-VQE grows
from about \(2.1\times\) to \(2{,}800\times\)
(Table~\ref{tab:n2_measurement_results}).  Including all three bond-breaking
pairs classically leaves a smaller quantum correction and reduces the
energy's sensitivity to its measured matrix elements.  In this comparison,
the improved reference primarily benefits sampling, whose cost also depends
on the operator variances.

\begin{table}[!ht]
\centering
\footnotesize
\setlength{\tabcolsep}{2pt}
\caption{Idealized N$_2$/STO--3G final-energy sampling costs at
\(\epsilon_{\mathrm{meas}}=1.6\mEh\), including occupation and seniority
killers.  For CASH-QSE,
\(|c_C|^2\) is the optimized classical-state weight; Q-SENSE-light has no
separate classical component.  A dash denotes a classical stop.  The final
column is ADAPT-VQE/method. Geometry and ADAPT baseline entries span all rows
at that geometry.}
\label{tab:n2_measurement_results}
\begin{ruledtabular}
\begin{tabular}{lccccc}
Partition & \shortstack{$R_{NN}$\\(\AA)} & $|c_C|^2$ & \shortstack{Method\\shots} & \shortstack{ADAPT\\shots} & \shortstack{ADAPT/\\method}\\
\hline
HF-CASH & \multirow{3}{*}{1.1} & 0.918 & $7.4\times10^4$ & \multirow{3}{*}{$3.3\times10^6$} & 45\\
CAS(4,4)-CASH &  & 0.975 & $2.3\times10^4$ &  & 150\\
CAS(6,6)-CASH &  & 0.991 & $2.6\times10^3$ &  & 1300\\
\midrule
HF-CASH & \multirow{3}{*}{1.8} & 0.426 & $1.1\times10^6$ & \multirow{3}{*}{$2.4\times10^6$} & 2.1\\
CAS(4,4)-CASH &  & 0.849 & $1.6\times10^5$ &  & 15\\
CAS(6,6)-CASH &  & 0.997 & $8.5\times10^2$ &  & 2800\\
\midrule
HF-CASH & \multirow{4}{*}{3.0} & 0.0718 & $2.3\times10^6$ & \multirow{4}{*}{$7.7\times10^5$} & 0.33\\
CAS(4,4)-CASH &  & 0.371 & $8.1\times10^5$ &  & 0.94\\
CAS(6,6)-CASH &  & 1.000 & -- &  & classical stop\\
Q-SENSE-light &  & -- & $2.2\times10^4$ &  & 35\\
\end{tabular}
\end{ruledtabular}
\end{table}

Coupling to external correlation can also require relaxation of the
active-space coefficients.
At 1.8~\AA{}, a frozen CAS(4,4) vector leaves a minimum error of
3.10~\(\mEh\), even when the complete external determinant space is
included.  Relaxing its coefficients in the hybrid Ritz problem at fixed
CASSCF orbitals restores chemical accuracy.  Enlarging the external space
alone therefore cannot repair the relative configuration weights of the
frozen classical component in this case.  The frozen-vector error floor and
coefficient diagnostics are reported in the Supplementary Information.

At dissociation, HF-CASH retains an approximately 18-fold complete-circuit
advantage over ADAPT-VQE but requires more shots
(Tables~\ref{tab:n2_circuit_results} and~\ref{tab:n2_measurement_results}).
CAS(4,4)-CASH also retains short circuits without a clear sampling advantage,
even after killer optimization.  The short preparations achieved by this
partition thus do not ensure inexpensive measurements.  Unlike the
dissociated H$_2$O results, the grouped measurements for these N$_2$
partitions remain costly, motivating a different organization of the quantum
components.

\subsection{Q-SENSE-light: recovering sampling efficiency through repartitioning}
\label{sec:qsense_light_results}

The dissociated N$_2$ results expose a limitation of using the
classical-reference/quantum-complement split as the first partition of Hilbert
space.  When the classical reference carries only a small fraction of the
target state and the structure of the resulting quantum complement does not
provide sufficient measurement reduction to compensate, CASH-QSE can retain
short preparation circuits while losing its sampling advantage.  This raises
a natural question: is the unfavorable measurement cost intrinsic to an
orthogonal subspace representation, or does it result from the order in which
the Hilbert space is partitioned?

To test the latter possibility, we first repartition the dissociated N$_2$
wave function by orbital-seniority patterns.  The resulting diagnostic
construction, which we call Q-SENSE-light and define in
Appendix~\ref{app:qsense_light}, represents the wave function by mutually
orthogonal states with fixed sets of singly occupied orbitals, without first
extracting a distinguished classical reference.  A classically tractable
component could subsequently be separated within any such seniority sector,
both to avoid remeasuring that component and, where useful, to refine the
remaining quantum state into shorter preparations.  The present benchmark
isolates the effect of changing the primary partition and does not apply this
additional within-sector classical boosting.

This seniority-first organization recovers a sampling advantage for
dissociated N$_2$: the resulting calculation requires about 35 times fewer
shots than ADAPT-VQE (Table~\ref{tab:n2_measurement_results}).  Its largest
preparation costs 208 CNOTs, compared with 99 for HF-CASH, and its largest
complete measurement circuit contains 470 CNOTs, including transition and
basis-change operations.  The sampling benefit therefore comes with longer
individual preparations; the comparison tests a change in both representation
and circuit budget.

Killers reduce the connected-only Q-SENSE-light cost by about
\(4.2\times\). A fixed-state control in the SI isolates the grouping
benefit for three dominant components with four singly occupied orbitals.  The occupation
constraints allow a regrouping of the measured terms that captures
cancellations within commuting fragments.  Keeping the original groups
removes the gain for these diagonal elements, while the variance of each
complete operator in the corresponding component is unchanged.  The remaining
sampling cost is dominated by transitions between components; the leading transitions
connecting the fully paired component to the four-singly-occupied
components are unaffected by the implemented killer operators.

For dissociated N$_2$, the comparison with HF-CASH therefore shows that the
order of partitioning is itself a resource-design choice: organizing the wave
function first by orbital seniority is more measurement-efficient here than
first extracting the HF component.  CAS(6,6) already reaches chemical accuracy
classically at this geometry, so Q-SENSE-light serves as a comparison of
alternative quantum representations rather than as the preferred treatment
of this particular system.  The present construction uses FCI projections to
select and compress the states; full Q-SENSE provides a route without that
projection shortcut \cite{Patel2026QSENSE}.

\section{Conclusions}
\label{sec:conclusions}

CASH-QSE is a hybrid quantum--classical method that evaluates the classical reference energy without quantum sampling while maintaining orthogonality among all components of the subspace expansion. It recovers correlation beyond a classical active-space reference by combining separately prepared quantum states through a classical eigenvalue problem. Its central idea is to use occupation structure both to organize the wave function and to simplify its measurement. This structure ensures orthogonality and supplies constraints for operator reductions that preserve the required matrix elements, supporting reductions in both state-preparation and measurement costs.

The H$_2$O and N$_2$ benchmarks reach chemical accuracy with short quantum circuits across weakly and strongly correlated geometries. The initial partition between classical and quantum contributions does not always yield quantum components that are simple to prepare. Further partitioning by fixed occupations and orbital seniority achieves short preparation circuits in the reported CASH-QSE benchmarks. These examples demonstrate that strong correlation in the total wave function can be represented through individually simple components. Including the relevant bond-breaking configurations in the classical active space can substantially reduce the sampling burden, but reference quality alone does not determine the benefit. Occupation constraints can also make measurements inexpensive when the classical component has little weight. Conversely, short preparation circuits can coexist with unfavorable sampling costs. The usefulness of a partition therefore depends on both the chemistry captured classically and the structure of the quantum correction. Repartitioning can improve sampling even when it increases the cost of individual preparations.

The demonstrated savings rely on FCI-assisted component selection and sampling estimates with known variances. Establishing practical efficiency requires accounting for the measurements used to select and optimize the states and to learn the identity shifts and fragment variances. The central next step is to identify favorable partitions from accessible chemical and structural information, and to determine whether their preparation and sampling benefits persist after these additional costs are included.

\begin{acknowledgments}
A.F.I. thanks Alexander Ibrahim, Alexey Uvarov, Praveen Jayakumar, and Davood Dar for useful comments and corrections. 
A.F.I. acknowledges financial support from the Natural Sciences and Engineering Research Council of Canada (NSERC). This research was enabled in part by computational resources provided by Compute Ontario (\url{https://computeontario.ca}) and the Digital Research Alliance of Canada (\url{https://alliancecan.ca}). Some computations were performed on the Niagara and Trillium supercomputers at the SciNet High Performance Computing Consortium and on the NARVAL and RORQUAL supercomputers through Calcul Qu\'ebec. SciNet is funded by Innovation, Science, and Economic Development Canada, the Digital Research Alliance of Canada, the Ontario Research Fund: Research Excellence, and the University of Toronto.
\end{acknowledgments}

\appendix

\section{Sector resolution and controllability}
\label{app:sector_control}

The CAS complement can be divided into orthogonal sectors at different
resolutions.  The choice determines both how many sectors are needed and
how restrictive their occupation conditions are for the unitary generators.
This appendix compares aggregate occupation conditions, complete occupation
patterns, and first defects defined by spin or spatial orbitals.  We ask
whether each choice admits a polynomial number of low-rank generators that
remain inside a sector and can prepare any state within it.  This comparison
motivates the spatial-orbital construction used in CASH-QSE and identifies
the assumptions needed for its state-preparation completeness.

Here \emph{state controllability}, or \emph{state completeness}, means that
products of the allowed unitaries can prepare any normalized real state in a
sector from a fixed reference, to arbitrary accuracy and up to global sign.  The
\emph{dynamical Lie algebra} is the real linear span of the generators and
their repeated commutators, restricted to that sector.  It need not contain
every real antisymmetric matrix for state completeness.  Below, fermionic
rank is the largest number of creation--annihilation pairs in any nonzero normal-ordered term, and
\(N_{\rm so}\) is the total number of spin orbitals.  A polynomial pool
size does not imply polynomial preparation length.

\paragraph{Aggregate occupation conditions.}
For inactive and external spatial-orbital sets \(\mathcal C,\mathcal V\),
with \(n_c=|\mathcal C|\), \(n_v=|\mathcal V|\), define
\[
 \hat h_c=2n_c I-\sum_{p\in\mathcal C,\sigma}\hat n_{p\sigma},
 \qquad
 \hat N_v=\sum_{p\in\mathcal V,\sigma}\hat n_{p\sigma},
\]
where \(\hat n_{p\sigma}=a_{p\sigma}^\dagger a_{p\sigma}\),
\(\sigma=\alpha,\beta\), and \(I\) is the identity.
Their eigenvalues \(h_c,N_v\) count inactive holes and external electrons.
At fixed total electron number, the CAS complement is \(h_c+N_v>0\),
or two sectors \(h_c>0\) and \(h_c=0,N_v>0\).
For example, the generators
\begin{equation}
 \left(1-\prod_{\substack{r\in\mathcal C,\,\tau=\alpha,\beta\\
                         r\tau\ne p\sigma}}\hat n_{r\tau}\right)
 \left(a_{q\sigma}^\dagger a_{p\sigma}
       -a_{p\sigma}^\dagger a_{q\sigma}\right),
 \label{eq:cash_aggregate_generator}
\end{equation}
with \(p\in\mathcal C\), \(q\notin\mathcal C\), preserve \(h_c>0\):
the prefactor permits the substitution only when another inactive hole
exists, so the last hole cannot be filled.  The factors commute, preserving
anti-Hermiticity, but their product can contain terms of fermionic rank
\(2n_c\).  Thus aggregate conditions admit explicit confined many-body
generators; confinement alone does not establish state completeness.

\paragraph{Fixed occupations and spin-orbital first defects.}
Resolving every inactive/external spin-orbital occupation gives up to
\(2^{2(n_c+n_v)}\) patterns.  Recording instead the first failed test in an
ordered list of \(n_{c\sigma}=1\), \(c\in\mathcal C\), and \(n_{v\sigma}=0\), \(v\in\mathcal V\), gives at most
\(2(n_c+n_v)\) complement sectors.  Both constructions fix individual
occupations.  Generalized singles and doubles acting only on the unfrozen
spin orbitals remain confined and provide an \(O(N_{\rm so}^4)\) pool for
the usual factorized unitary coupled-cluster state-completeness construction
\cite{Evangelista2019ExactUCC}.  Further symmetry restrictions require
separate justification; individual spin-orbital conditions need not
preserve total spin.

\paragraph{Spatial-orbital first defects.}
The construction used here gives at most \(n_c+n_v\) complement sectors.
Orbitals preceding the first defect are fixed to their CAS occupations.  At the defective spatial orbital
\(p\), the allowed occupations are \(0,\alpha,\beta\) for a core defect
and \(\alpha,\beta,\alpha\beta\) for an external defect.  Within these
respective spaces, their projectors are
\begin{align}
 (\hat P_0^c,\hat P_\alpha^c,\hat P_\beta^c)
 &=(1-\hat n_{p\alpha}-\hat n_{p\beta},
       \hat n_{p\alpha},\hat n_{p\beta}),\notag\\
 (\hat P_\alpha^v,\hat P_\beta^v,\hat P_{\alpha\beta}^v)
 &=(1-\hat n_{p\beta},1-\hat n_{p\alpha},
       \hat n_{p\alpha}+\hat n_{p\beta}-1).
 \label{eq:cash_local_selectors}
\end{align}
These expressions are used only within the stated defect space.
Let \(R\) contain the unfrozen spin orbitals other than
\(p\alpha,p\beta\), and let \(\mathcal P_R\) be the generalized
singles/doubles pool on \(R\), preserving spin-resolved particle numbers:
\[
 \hat\kappa_j^i=a_i^\dagger a_j-\mathrm{h.c.},
 \qquad
 \hat\kappa_{kl}^{ij}=a_i^\dagger a_j^\dagger a_l a_k-\mathrm{h.c.}
\]
Here \(i,j,k,l\in R\); indices may overlap between the creation and annihilation pairs, and
\(\mathrm{h.c.}\) means the Hermitian conjugate.  Multiplication by the
appropriate projector \(\hat P_\eta\) from
Eq.~\eqref{eq:cash_local_selectors} confines a generator to local
occupation \(\eta\), with rank at most three.
The following rank-two generators change the occupation of \(p\):
\begin{align}
 \hat B^c_{p\sigma,q\sigma}
 &=(1-\hat n_{p\bar\sigma})
 (a_{p\sigma}^\dagger a_{q\sigma}-\mathrm{h.c.}),\notag\\
 \hat B^v_{p\sigma,q\sigma}
 &=\hat n_{p\bar\sigma}
 (a_{p\sigma}^\dagger a_{q\sigma}-\mathrm{h.c.}),
 \label{eq:cash_local_bridges}
\end{align}
where \(q\sigma\in R\) and \(\bar\sigma\) is the opposite spin.
The prefactors prevent double occupation for a core defect and empty
occupation for an external defect.  Denote their nonzero members for the
chosen defect type by \(\mathcal B\).  An explicit rank-three pool is
\begin{equation}
 \mathcal G_3=
 \{\hat P_\eta\hat\kappa:\hat\kappa\in\mathcal P_R,
       \ \eta\text{ allowed}\}\cup\mathcal B.
 \label{eq:cash_rank_three_pool}
\end{equation}

Assume that \(\mathcal P_R\) is state-complete within each nonempty
local-occupation subspace and that the displayed generators connect each
singly occupied subspace to the empty one (core defect) or doubly occupied
one (external defect).  Then \(\mathcal G_3\) is state-complete.
For a core defect, the projected residual generators first rotate the
components with occupations \(0\) and \(\sigma\) into two determinants
differing by one \(q\sigma\to p\sigma\) substitution.  Exponentiating
\(\hat B^c_{p\sigma,q\sigma}\) transfers their combined amplitude into
the empty-\(p\) subspace without changing the opposite-spin subspace.
Repeating for both spins, followed by a residual transformation, maps any
state to a fixed reference.  Reversing this sequence prepares the target.
The external case follows by exchanging particles and holes.  A single
nonempty subspace needs only its residual pool.

Rank three is unnecessary when suitable auxiliary spin orbitals exist.
Consider the rank-two pool
\begin{equation}
 \mathcal G_2=\mathcal P_R\cup\mathcal B\cup
 \{\hat n_{p\sigma}\hat\kappa_j^i:
       \hat\kappa_j^i\in\mathcal P_R,\ \sigma=\alpha,\beta\}.
 \label{eq:cash_rank_two_pool}
\end{equation}
The added terms are already generalized doubles:
\begin{equation}
 \hat n_{p\sigma}\hat\kappa_j^i
 =a_{p\sigma}^\dagger a_i^\dagger a_j a_{p\sigma}-\mathrm{h.c.}
 \label{eq:cash_overlap_double}
\end{equation}
The needed rank-three terms follow from
\begin{align}
 [\hat n_{p\sigma}\hat\kappa_m^i,\hat\kappa_{kl}^{mj}]
 &=\hat n_{p\sigma}\hat\kappa_{kl}^{ij},
 \label{eq:cash_controlled_double_comm}\\
 [\hat n_{p\sigma}\hat\kappa_m^i,\hat n_k\hat\kappa_j^m]
 &=\hat n_{p\sigma}\hat n_k\hat\kappa_j^i.
 \label{eq:cash_overlap_control_comm}
\end{align}
In the first identity, \(i,j,k,l,m\in R\) are distinct and \(m,i\)
have the same spin; in the second, \(i,j,k,m\in R\) are distinct and
\(i,j,m\) have the same spin.  Here \(\hat n_k=a_k^\dagger a_k\).
All substitutions must preserve the imposed symmetries.  Together with
Eq.~\eqref{eq:cash_local_selectors}, these identities recover
\(\mathcal G_3\) from commutators and linear combinations of
\(\mathcal G_2\).  Thus an \(O(N_{\rm so}^4)\) rank-two pool suffices
under the stated residual-completeness and auxiliary-orbital conditions.
If auxiliary orbitals are unavailable, or additional symmetries restrict
the generators, a separate argument is required.

\section{Orbital relaxation and the self-consistent macrocycle}
\label{app:orbital_macrocycle}

This formal extension is not used to generate the reported resource ratios.

The CAS-coefficient update alone does not make the total energy stationary
with respect to the orbital partition.  Orbital relaxation is implemented by
the unitary orbital-rotation operator
\begin{align*}
  \hat U_{\mathrm{orb}}(\vct\kappa)
  &=\exp\!\left[\hat K(\vct\kappa)\right],\\
  \hat K(\vct\kappa)
  &=\sum_{(p,q)\in\mathcal R_{\rm orb}}
  \kappa_{pq}(\hat E_{pq}-\hat E_{qp}),\\
  \hat E_{pq}&=\sum_\sigma a_{p\sigma}^\dagger a_{q\sigma},
\end{align*}
Here \(\vct\kappa\) collects the real orbital-rotation parameters \(\kappa_{pq}\). The one-body generator \(\hat K\) satisfies \(\hat K^\dagger=-\hat K\), and
\(\mathcal R_{\rm orb}\) contains symmetry-allowed nonredundant orbital pairs.
For CASSCF, the nonredundant orbital rotations change the relative inactive,
active, and external subspaces.  Rotations wholly within the inactive or external spaces
leave the reference invariant, while active--active one-electron rotations
are redundant under complete reoptimization of the CAS coefficients.  This
redundancy concerns the classical orbital parameterization only; generalized
fermionic single and double generators with active indices may still appear
in the quantum-state ansatz when they preserve the assigned structured
subspace.

With
\(\Hhat(\vct\kappa)
=\hat U_{\mathrm{orb}}^\dagger\Hhat\hat U_{\mathrm{orb}}\),
the orbital derivative of the optimized Ritz root is
\begin{equation}
  \frac{\partial E}{\partial\kappa_{pq}}
  =\vct c^\dagger
  \frac{\partial\vct H_{\mathrm{hyb}}}{\partial\kappa_{pq}}
  \vct c.
  \label{eq:hybrid_orbital_gradient}
\end{equation}
At \(\vct\kappa=0\),
\(\partial\Hhat/\partial\kappa_{pq}
=[\Hhat,\hat E_{pq}-\hat E_{qp}]\).
The orbital step can therefore be obtained with a conventional
augmented-Hessian, quasi-Newton, or trust-region CASSCF optimizer.

The complete self-consistent macrocycle alternates the coupled variational
degrees of freedom:
\begin{enumerate}
  \setlength{\itemsep}{0.35em}
  \item The orbitals and CAS vector are initialized with a conventional CASSCF
  calculation.
  \item The first-defect and refined structured subspaces are defined in that
  orbital basis.
  \item The retained quantum states are optimized at fixed CAS coefficients and
  orbitals using the global Ritz objective.
  \item The CAS coefficients are updated with
  Eq.~\eqref{eq:uncontracted_problem}, and the contracted state is normalized
  with Eq.~\eqref{eq:normalized_cas_update}.
  \item The orbital gradient in Eq.~\eqref{eq:hybrid_orbital_gradient} is evaluated,
  an allowed orbital step is taken, the integrals are transformed, and the operators
  defining the structural subspaces are rebuilt in the new orbital basis.
  \item The quantum states are warm-started by retaining their occupation
  labels, seed coefficients, generator index sequences, and circuit parameters,
  interpreted in the updated orbital basis.  The circuit parameters are then
  reoptimized.  The macrocycle repeats until the energy, CAS state, orbital
  gradient, and circuit-gradient criteria are simultaneously satisfied.
\end{enumerate}
At convergence, the CASH-QSE energy is stationary with respect to the CAS
coefficients, the nonredundant CASSCF orbital rotations, and the retained
quantum-circuit parameters.  The classical/quantum basis remains orthogonal in
the current orbital basis.

\section{Supplementary computational specifications}
\label{app:computational_specs}

\subsection{Orbital partitions and tracking}
\label{app:orbital_specs}

For H$_2$O, the molecule lies in the \(yz\) plane with its \(C_2\) axis
along \(z\).  The STO--3G CASSCF(2,2) orbital partition is
\begin{align*}
  \mathcal C_{\mathrm{H_2O}}&=\{1a_1,2a_1,1b_2,1b_1\},\\
  \mathcal A_{\mathrm{H_2O}}&=\{3a_1,4a_1\},\\
  \mathcal V_{\mathrm{H_2O}}&=\{2b_2\}.
\end{align*}
The CAS(4,4) diagnostic adds the complementary bond-breaking pair, giving
the active set \(\{3a_1,1b_2,4a_1,2b_2\}\).

For N$_2$/STO--3G, the CASSCF(6,6) partition is
\begin{align*}
 \mathcal C_{\mathrm{N_2}}&=\{1\sigma_g,1\sigma_u,2\sigma_g,2\sigma_u\},\\
 \mathcal A_{\mathrm{N_2}}&=\{1\pi_{u,x},1\pi_{u,y},3\sigma_g,
 1\pi^*_{g,x},1\pi^*_{g,y},3\sigma^*_u\},\\
 \mathcal V_{\mathrm{N_2}}&=\varnothing.
\end{align*}
The CASSCF(4,4) partition retains the complete \(\pi/\pi^*\) manifold,
\begin{align*}
 \mathcal C_{44}&=\{1\sigma_g,1\sigma_u,2\sigma_g,2\sigma_u,3\sigma_g\},\\
 \mathcal A_{44}&=\{1\pi_{u,x},1\pi_{u,y},1\pi^*_{g,x},1\pi^*_{g,y}\},\\
 \mathcal V_{44}&=\{3\sigma^*_u\}.
\end{align*}
The CAS(2,2) overlap diagnostic uses the
\(3\sigma_g/3\sigma_u^*\) pair.

After freezing the \(1a_1\) core orbital, the ordered H$_2$O/cc-pVDZ
space is
\begin{equation}
\begin{split}
\mathcal W_{14}={}&(2a_1,1b_1;\ 1b_2,3a_1,4a_1,2b_2;\\
&5a_1,3b_2,6a_1,2b_1,4b_2,7a_1,1a_2,3b_1),
\end{split}
\label{eq:h2o14_window}
\end{equation}
The semicolons separate the retained inactive orbitals, the CAS(4,4)
active orbitals, and the eight external virtual orbitals.
The equilibrium orbital identities in Eq.~\eqref{eq:h2o14_window} are
tracked to 1.75 and 3.00~\AA{} by maximum overlap within each irreducible
representation of \(C_{2v}\).  They are never re-sorted by the instantaneous
RHF orbital energies.

\subsection{Parity operators and tapering sectors}
\label{app:parity_specs}

For a set \(G\) of spin-orbital indices, Jordan--Wigner mapping gives
\[
(-1)^{\sum_{i\in G}\hat n_i}=\prod_{i\in G}Z_i.
\]
Here \(Z_i\) is the Pauli \(Z\) operator on qubit \(i\). The spin parities are \(\hat P_\alpha=(-1)^{\hat N_\alpha}\) and \(\hat P_\beta=(-1)^{\hat N_\beta}\), with \(\hat N_\sigma=\sum_p\hat n_{p\sigma}\).

For H$_2$O/STO--3G, \(N_\alpha=N_\beta=5\) and the target irreducible
representation is \(A_1\) of \(C_{2v}\simeq\mathbb Z_2^2\).
The independent spatial occupation parities are
\(\hat Q_{B_1}=(-1)^{\hat N_{B_1}}\) and
\(\hat Q_{B_2}=(-1)^{\hat N_{B_2}}\), where \(\hat N_\Gamma\) counts
electrons of both spins in orbitals belonging to irreducible representation
\(\Gamma\).  The ordered set
\(\{\hat P_\alpha,\hat P_\beta,\hat Q_{B_1},\hat Q_{B_2}\}\)
contains four independent commuting Pauli-product symmetries with target
eigenvalues \((-1,-1,+1,+1)\).
Any CASSCF orbital rotations are restricted to individual
\(A_1,A_2,B_1,B_2\) blocks.
For H$_2$O/cc-pVDZ, the retained \(A_2\) orbital contributes to both
independent spatial parities:
\[
 \hat Q_1=(-1)^{\hat N_{B_1}+\hat N_{A_2}},\qquad
 \hat Q_2=(-1)^{\hat N_{B_2}+\hat N_{A_2}}.
\]
Together with \(\hat P_\alpha\) and \(\hat P_\beta\), these give four
generators with eigenvalues \((+1,+1,+1,+1)\) in the
\(N_\alpha=N_\beta=4\), \(A_1\) sector. These are the operators used by
the implementation, giving the stated 28-to-24-qubit reduction.

For N$_2$/STO--3G, \(N_\alpha=N_\beta=7\) and the target irreducible
representation is \(A_g\) of \(D_{2h}\simeq\mathbb Z_2^3\).
The spatial parities are
\(\hat Q_{\pi_x}=(-1)^{\hat N_{\pi_x}}\),
\(\hat Q_{\pi_y}=(-1)^{\hat N_{\pi_y}}\), and
\(\hat Q_u=(-1)^{\hat N_u}\).
Here \(\hat N_{\pi_x}\) and \(\hat N_{\pi_y}\) count electrons in both
gerade and ungerade orbitals of the indicated Cartesian \(\pi\) component,
while \(\hat N_u\) counts electrons in all ungerade orbitals.
The ordered set
\(\{\hat P_\alpha,\hat P_\beta,\hat Q_{\pi_x},\hat Q_{\pi_y},\hat Q_u\}\)
contains five independent commuting Pauli-product symmetries with target
eigenvalues \((-1,-1,+1,+1,+1)\).
Orbital rotations remain block diagonal within the irreducible
representations of \(D_{2h}\).

\subsection{State-preparation specifications}
\label{app:state_preparation_specs}

For H$_2$O/STO--3G, HF has five nonempty occupied-orbital defect sectors
and no virtual-first sector: fixing all five occupied orbitals already
accounts for all ten electrons. CAS(2,2) has four inactive-orbital defect
sectors and one nonempty virtual-first sector. Both constructions therefore
have five broad first-defect states.
For N$_2$/STO--3G, CAS(6,6) has four first-inactive-defect sectors,
while HF can resolve its seven occupied orbitals separately.

In the direct construction of the final N$_2$/STO--3G and
H$_2$O/cc-pVDZ bases, seniority-zero blocks use pair rotations;
seniority-two blocks use singlet-CSF seeds followed by pair rotations.
Higher-seniority blocks use compact valence-bond or open-shell singlets
with the required pair or spin-recoupling rotations.
The H$_2$O CAS(2,2) classical reference has a coherent 64-CNOT preparation.
The N$_2$ CAS(4,4) and CAS(6,6) coherent preparations exceed the 100-CNOT
cap; their orthogonal components with fixed open-shell and paired-occupation
patterns each require at most 96 CNOTs.

\subsection{Circuit checks and numerical records}
\label{app:implementation_checks}

The ten-qubit H$_2$O/STO--3G extended-swap implementation contains a
70-CNOT controlled-SWAP network before the FC-SI basis change.
The corresponding 15-qubit N$_2$/STO--3G network contains 105 CNOTs.

Statevector checks verify independence and mutual commutation of the
symmetry generators, their target eigenvalues on all basis states, and
agreement of tapered and untapered Ritz energies and transition elements.
For each transition circuit they also verify logical-idle-state invariance,
the ancilla estimator including relative phases, and idle-register
disentanglement. Under inverse tapering, logical zero corresponds to the
determinant with the lowest spatial orbital doubly occupied for both
STO--3G systems. Every retained unconditional generator leaves this pair
untouched and annihilates that determinant. For cc-pVDZ, whose target
parities are all positive, logical zero maps to the physical vacuum.
The SI contains the per-preparation audit; no additional gates are required.

The Supplementary Information contains exact input geometries, unrounded
numerical values, and the records needed to reconstruct the calculations.
The electronic-structure records include determinant dimensions, orbital
tracking overlaps, and solver diagnostics.  The state-construction records
include seeds, tie-breaking decisions, classical-component weights,
generator sequences, transition-reconstruction errors, and symmetry checks.
The circuit records include Clifford maps, removed qubits, and tapered
generator pools.  ADAPT-VQE search records and convergence diagnostics
document the comparator selection.

\section{\texorpdfstring{Discrete sampling allocations for H$_2$O/cc-pVDZ}{Discrete sampling allocations for H2O/cc-pVDZ}}
\label{app:discrete_allocations}

Table~\ref{tab:discrete_allocations} records the one-shot-per-group check for
the 14-orbital CAS(4,4) benchmark.  The floor applies to both methods; exact
fragment variances are assumed and pilot-shot overhead is excluded. At
1.75~\AA{}, the accepted ADAPT-VQE state requires 20,709,831 shots after
rounding with a one-shot minimum per group. The 3.0~\AA{} ADAPT allocation
uses the FCI proxy specified in Computational details.

\begin{table*}[t]
\caption{Continuous and one-shot-floor allocations for H$_2$O/cc-pVDZ.
The ratio is ADAPT-VQE/CASH-QSE with the same floor applied to both methods.}
\label{tab:discrete_allocations}
\begin{ruledtabular}
\begin{tabular}{ccccc}
$R_{\rm OH}$ (\AA)&CASH FC-SI groups&CASH continuous shots&CASH floor shots&Floor ratio\\
\hline
0.96&33,856&$2.1\times10^4$&$5.18\times10^4$&560\\
1.75&23,073&$1.2\times10^4$&$3.28\times10^4$&630\\
3.0&486&$5.1\times10^2$&$9.41\times10^2$&$4.4\times10^3$\\
\end{tabular}
\end{ruledtabular}
\end{table*}

\section{Q-SENSE-light seniority repartitioning diagnostic}
\label{app:qsense_light}

Here \(\mathcal S\) denotes a set of singly occupied spatial orbitals, and
\(\hat{\mathcal P}_{\mathcal S}\) projects onto its fixed orbital-seniority pattern.  For the normalized
FCI state and patterns with \(w_{\mathcal S}>0\), we define
\begin{equation*}
 |C_{\mathcal S}\rangle=
 \frac{\hat{\mathcal P}_{\mathcal S}|\Psi_{\rm FCI}\rangle}
 {\sqrt{w_{\mathcal S}}},
 \qquad
 w_{\mathcal S}=
 \langle\Psi_{\rm FCI}|\hat{\mathcal P}_{\mathcal S}|\Psi_{\rm FCI}\rangle.
\end{equation*}
The nonzero \(|C_{\mathcal S}\rangle\) are orthonormal because distinct seniority
patterns have disjoint determinant support.
Q-SENSE-light orders them by decreasing \(w_{\mathcal S}\) and retains the
smallest collection \(\mathcal R_{\rm ret}\) satisfying
\begin{equation*}
 \sum_{\mathcal S\in\mathcal R_{\rm ret}}w_{\mathcal S}\ge 0.9999.
\end{equation*}
For N$_2$/STO--3G at 3.0~\AA{}, \(|\mathcal R_{\rm ret}|=11\) and the retained weight is approximately
0.9999.

For each pattern, \(\hat{\mathcal P}_{\pi}\) fixes one paired-orbital
occupation, and we choose
\(\pi_{\mathcal S}^*=\arg\max_{\pi}
\|\hat{\mathcal P}_{\pi}|C_{\mathcal S}\rangle\|^2\).  The seed
\begin{equation*}
 |\chi_{\mathcal S}\rangle=
 \frac{\hat{\mathcal P}_{\pi_{\mathcal S}^*}|C_{\mathcal S}\rangle}
 {\|\hat{\mathcal P}_{\pi_{\mathcal S}^*}|C_{\mathcal S}\rangle\|}
\end{equation*}
is a spin-singlet CSF fixing all unpaired orbitals.  ADAPT restricted to
pair rotations then uses
\begin{equation*}
 \hat U_{pq}(\theta)=\exp\!\left[\theta
 (\hat P_p^\dagger \hat P_q-\hat P_q^\dagger \hat P_p)\right],
 \qquad \hat P_p^\dagger=a_{p\alpha}^\dagger a_{p\beta}^\dagger,
\end{equation*}
with \(p,q\notin\mathcal S\).
At each iteration, the unitary exponential of the generator with the largest
absolute fidelity gradient is appended to the circuit, and all angles are
reoptimized to minimize
\(1-|\langle C_{\mathcal S}|\hat U_{\mathcal S}|
\chi_{\mathcal S}\rangle|^2\), where \(\hat U_{\mathcal S}\) is the product of accepted pair rotations. Repeated generators are allowed.
Each pair generator is exponentiated exactly because its mapped Pauli words
commute; no product formula is used.  The retained circuit is the first point
with fidelity at least 0.9999.  A Schmidt-decomposition check between
paired occupations and spin assignments places the one-CSF fidelity ceiling well above the 0.9999 target,
so the seed restriction is negligible at this threshold.

Let \(\ket{\widetilde C_{\mathcal S}}=\hat U_{\mathcal S}\ket{\chi_{\mathcal S}}\). These approximate states remain orthogonal because their singly occupied sets are unchanged.  We diagonalize
\(H_{\mathcal S\mathcal T}=
\langle\widetilde C_{\mathcal S}|\Hhat|
\widetilde C_{\mathcal T}\rangle\) in the 11-state
basis.  Every nonzero diagonal and off-diagonal element is rebuilt with the
same connected-term filter, killer operators from left- and right-state
occupation and local seniority constraints, FC-SI grouping, exact fragment
variances, and continuous optimal allocation as in the CASH-QSE calculations.
This construction intentionally uses
FCI projections to isolate the effect of repartitioning.  It is therefore
``Q-SENSE-light,'' not a from-scratch solution of the full Q-SENSE equations;
the latter is given in Ref.~\cite{Patel2026QSENSE}.

\end{document}